\documentclass[a4paper,11pt]{article}

\usepackage{amsmath,amssymb,amsfonts,amsthm}
\usepackage[titletoc,title]{appendix}

\numberwithin{equation}{section}

\newcommand{\bea}{\begin{eqnarray}}
\newcommand{\eea}{\end{eqnarray}}
\newcommand{\be}{\begin{equation}}
\newcommand{\ee}{\end{equation}}
\newcommand{\nn}{\nonumber \\}

\begin{document}
\begin{titlepage}

\vfill \vfill \vfill
\begin{center}
{\bf\LARGE Deformed ${\cal N}$\,=\,8 mechanics of ${\bf (8, 8, 0)}$ multiplets}
\end{center}
\vspace{1.5cm}

\begin{center}
{\bf\Large Evgeny Ivanov${\,}^{a)}$, Olaf Lechtenfeld${\,}^{b)}$, Stepan Sidorov${\,}^{a)}$}\\
\end{center}
\vspace{0.4cm}

\centerline{${\,}^{a)}$ \it Bogoliubov Laboratory of Theoretical Physics, JINR, 141980 Dubna, Moscow Region, Russia}
\vspace{0.2cm}

\centerline{${\,}^{b)}$ \it Institut f\"ur Theoretische Physik
and Riemann Center for Geometry and Physics,}
\centerline
{\it Leibniz Universit\"at Hannover,
Appelstra{\ss}e 2, 30167 Hannover, Germany.}
\vspace{0.3cm}

\centerline{eivanov@theor.jinr.ru,~lechtenf@itp.uni-hannover.de,~sidorovstepan88@gmail.com}
\vspace{0.2cm}

\vspace{2cm}

\par
\begin{center}
{\bf ABSTRACT}
\end{center}

\noindent
We construct new models of ``curved'' SU$(4|1)$ supersymmetric mechanics  based on
two versions of the off-shell multiplet ${\bf (8, 8, 0)}$ which are ``mirror'' to each other.
The worldline realizations of the supergroup SU$(4|1)$ are treated as a deformation
of flat ${\cal N}\,{=}\,8$, $d\,{=}\,1$ supersymmetry. Using SU$(4|1)$ chiral superfields,
we derive invariant actions for the first-type ${\bf (8, 8, 0)}$ multiplet, which parametrizes
special K\"ahler manifolds. Since we are not aware of a manifestly SU$(4|1)$ covariant
superfield formalism for  the second-type ${\bf (8, 8, 0)}$ multiplet, we perform a general
construction of SU$(4|1)$ invariant actions for both multiplet types in terms of SU$(2|1)$
superfields. An important class of such actions enjoys superconformal OSp$(8|2)$ invariance.
We also build off-shell actions for the SU$(4|1)$ multiplets ${\bf (6, 8, 2)}$ and ${\bf (7, 8, 1)}$
through appropriate substitutions for the component fields in the ${\bf (8, 8, 0)}$ actions.
The ${\bf (6, 8, 2)}$ actions are shown to respect superconformal SU$(4|1,1)$ invariance.

\vfill{}


\end{titlepage}

\section{Introduction}
In recent years, mainly motivated by the study of higher-dimensional models with  ``curved'' rigid supersymmetries (see e.g.~\cite{FS}), there was a growth of activity  in
supersymmetric mechanics (SM) models underlain by some semi-simple  superalgebras treated as deformations of flat one-dimensional supersymmetries with
the same number of supercharges. The simplest superalgebra  of this kind is $su(2|1)$ (and its central-charge extension $\widehat{su}(2|1)$), which is  a deformation
of rigid ${\cal N}\,{=}\,4, d\,{=}\,1$ supersymmetry by a mass-dimension parameter $m$. The first examples with a worldline realization of  $su(2|1)$ supersymmetry were considered
more than 10 years ago (prior to  \cite{FS} and related works) in
\cite{BelNer1},  \cite{R1} and in \cite{WS} (where it was  named ``weak  $d\,{=}\,1$ supersymmetry'').
The corresponding worldline $su(2|1)$  multiplets had $d\,{=}\,1$ field contents ${\bf (2, 4, 2)}$ and ${\bf (1, 4, 3)}$.\footnote{Our notation follows ref.~\cite{ToPash}:
bold numerals denote, respectively, the number of physical bosonic, physical fermionic and auxiliary bosonic degrees of freedom in the given supermultiplet.}

A systematic superfield approach to $su(2|1)$
supersymmetry was  worked out in \cite{DSQM}, \cite{SKO}, \cite{ISTconf} and \cite{DHSS}. The models built on the multiplets ${\bf ( 1, 4, 3)}$, ${\bf (2, 4, 2)}$ and ${\bf (4, 4, 0)}$
were studied at the classical and quantum level. Recently, $su(2|1)$ invariant versions of super Calogero-Moser systems
were constructed and quantized \cite{FI16}, \cite{FeI}, \cite{FeILeSid}. The common notable features of all these models are:
\begin{itemize}
\item Oscillator-type Lagrangians for the bosonic fields, with $m^2$ as the oscillator strength,
\item Wess-Zumino type terms for the bosonic fields, of the type $\sim i m(\dot z \bar z - z \dot{\bar z})$,
\item At the lowest energy levels, wave functions form {\it atypical}  $su(2|1)$ multiplets,
with unequal numbers of the bosonic and fermionic states.
\end{itemize}

It was of obvious interest to move one step further and to consider mechanics models with analogous deformations of  ${\cal N}\,{=}\,8, d\,{=}\,1$
supersymmetry. In contrast to ${\cal N}\,{=}\,4$ supersymmetry, in the ${\cal N}\,{=}\,8$ case there exist two different possibilities for deformation
due to the existence of two different superalgebras with eight supercharges: $su(2|2)$ and $su(4|1)$, with $R$-symmetry algebra $su(2)\oplus su(2)$
or $su(4) \oplus u(1)$, respectively.\footnote{In the $su(2|2)$  case one can also add two central charges.}
The $su(2|2)$ models have been considered in \cite{su22} by analogy with the $su(2|1)$ case, on the basis of the appropriate  superfield worldline
formalism, as deformations of flat ${\cal N}\,{=}\,8$ SM models \cite{BIKL1}, \cite{BIKL2}, \cite{BHSS}, \cite{ILS}, \cite{286}.
They were built on the off-shell multiplets $({\bf 3, 8, 5})$, $({\bf 4, 8, 4})$ and $({\bf 5, 8, 3})$.
One class of $({\bf 5, 8, 3})$ actions represents a massive deformation for the same multiplet  in the flat case \cite{IvSmi}, \cite{DR}.
Another class enjoys superconformal ${\rm OSp}(4^*|4)$ invariance. Remarkably, the superconformal group ${\rm OSp}(4^*|4)$ is a closure of its two different ${\rm SU}(2|2)$ subgroups,
with deformation parameters $m$ and $-m$. So any ${\rm SU}(2|2)$ invariant action involving only even powers of $m$ is automatically
superconformal. Based on this observation, the general ${\rm SU}(2|2)$ action of  the multiplet $({\bf 3, 8, 5})$ was shown to be superconformal.

It turns out that some admissible multiplets of flat ${\cal N}\,{=}\,8$ supersymmetry do not have ${\rm SU}(2|2)$ analogs, most
importantly the so called ``root'' ${\cal N}\,{=}\,8$ multiplet $({\bf 8, 8, 0})$. The significance of this root multiplet derives from the fact that all
other flat ${\cal N}\,{=}\,8$ multiplets
and their invariant actions can be obtained from the root one and its general actions through appropriate covariant substitution of the auxiliary fields
(or Hamiltonian reductions, in the Hamiltonian
formalism) \cite{ILS} as a generalization of the phenomenon found in \cite{ToPash} at the linearized level.\footnote{As an aside, the multiplet $({\bf 8, 8, 0})$
has a puzzling relationship with the
octonion algebra \cite{FTo}.}
Deforming the flat (${\bf 8, 8, 0}$) multiplet has remained an open problem.

In the present paper we show that the latter becomes possible within the alternative ${\rm SU}(4|1)$ deformation. Interestingly, there exist
two such root ${\rm SU}(4|1)$ multiplets, which are  complementary
to each other in the sense that the ${\rm SU}(4)$ assignments of their fermionic and bosonic components are interchanged. Namely, in one multiplet, the bosonic $d{=}1$ fields
are in $\underline{\bf 1} \oplus \underline{\bf 1}^{*}  \oplus \underline{\bf 6}$ of ${\rm SU}(4)$ (eight real fields)  and the fermionic fields
in $\underline{\bf 4}\oplus \underline{\bf 4}^{*}$ (4 complex fields),
while in the other multiplet the
bosonic fields are in $\underline{\bf 4} \oplus \underline{\bf 4}^{*}$ and the fermionic fields in
$\underline{\bf 1} \oplus \underline{\bf 1}^{*} \oplus \underline{\bf 6}$\,.
In the ``flat'' ${\cal N}\,{=}\,8, d\,{=}\,1$ limit they go over to two different 8-dimensional multiplets of the SO(8) $R$-symmetry related by
triality (see, e.g., \cite{GSW1}, \cite{GSW2}).
These two multiplets are analogs of the mutually
``mirror'' ${\cal N}\,{=}\,4$ multiplets  $({\bf 4, 4, 0)}$, for which bosonic and fermionic components form doublets with respect to different ${\rm SU}(2)$ factors of the SO(4)
$R$-symmetry. For this reason it is natural to treat the two root ${\rm SU}(4|1)$ ${\bf (8, 8, 0)}$ multiplets as ``mirror'' to each other.

The main incentive  of our paper is constructing invariant actions for both types of the ${\bf (8, 8, 0)}$ multiplets.  To this end, we will use a manifestly ${\rm SU}(4|1)$ covariant
superspace formalism along with the ${\rm SU}(2|1)$ superfield approach, in which the extra ${\rm SU}(4|1)/{\rm SU}(2|1)$ transformations are realized in a hidden way.
In some cases, it is simplest to use the component approach.  The point
is that ${\rm SU}(4|1)$ possesses many non-equivalent worldline
supercosets, including the  harmonic ones \cite{HSS}, and it is not easy to decide which superfield formalism is most
adequate for one or another ${\rm SU}(4|1)$ multiplet. We utilize several versions of such an extended superfield approach
for constructing invariant actions.

The paper is organized as follows. In Section~2 we present the superalgebra $su(2|1)$ and describe the  relevant worldline supercosets. In Section~3, on the example of flat
${\cal N}\,{=}\,8, d\,{=}\,1$ supersymmetry, we discuss  three possible $({\bf 8, 8, 0)}$ multiplets, which are not equivalent if the ${\rm SO}(8)$ $R$-symmetry is broken, and argue that only two
of them can be extended to the deformed ${\rm SU}(4|1)$ case. The various superfield and component descriptions of the first version of the ${\rm SU}(4|1)$ $({\bf 8, 8, 0)}$ multiplet
are the subject of Section~4. We find three different classes of invariant actions for this multiplet, including an ${\rm OSp}(8|2)$ invariant one,
with an $R$-symmetry enhanced to ${\rm SO}(8)$. The analogous treatment of the second version of the multiplet $({\bf 8, 8, 0)}$ is given in Section~5. We show that its general
invariant action is superconformal and equivalent to the superconformal action of the first version. Summary and outlook are given in Section~6. An Appendix~A contains
details of calculating the invariant actions in the appropriate harmonic ${\rm SU}(4|1)$ superspaces, and in Appendices B and~C the off-shell actions for the ${\rm SU}(4|1)$
multiplets $({\bf 6, 8, 2})$ and $({\bf 7, 8, 1})$ are presented. The full set of (anti)commutation relations of the conformal superalgebra $osp(8|2)$ is given in Appendix~D.

\section{Supergroup  ${\rm SU}(4|1)$ and its worldline realizations}
We consider ${\rm SU}(4|1)$ supersymmetry as a deformation of the standard ${\cal N}\,{=}\,8$, $d\,{=}\,1$ supersymmetry \cite{BIKL1}, \cite{BIKL2}, \cite{BHSS}, \cite{ILS}.
The superalgebra $su(4|1)$ is given by the following non-vanishing (anti)commutators:
\bea
    &&\left\lbrace Q^{I}, \bar{Q}_{J}\right\rbrace = 2m\,L^{I}_{J} + 2\delta^{I}_{J}{\cal H},\qquad
    \left[L^I_J,  L^K_L\right]
    = \delta^K_J L^I_L - \delta^I_L L^K_J,\nn
    &&\left[L^I_J, Q^{K}\right]
    = \delta^K_J Q^{I} - \frac{1}{4}\,\delta^I_J Q^{K} ,\qquad
    \left[L^I_J, \bar{Q}_{L}\right] = \frac{1}{4}\,\delta^I_J\bar{Q}_{L}-\delta^I_L\bar{Q}_{J}\,,\nn
    &&\left[{\cal H}, Q^{K}\right]=-\,\frac{3m}{4}\,Q^{K},\qquad
    \left[{\cal H}, \bar{Q}_{L}\right]=\frac{3m}{4}\,\bar{Q}_{L}\,.\label{algebra_su41}
\eea
Here, $L^I_J$ are the generators of the $R$-symmetry group  ${\rm SU}(4)$, and the capital indices $I,J,K,L$ ($I\,{=}\,1,2,3,4$) refer
to the ${\rm SU}(4)$ fundamental (``quark'') representation and its conjugate. ${\cal H}$ is the U(1) generator. In the contraction limit $m=0$ the above superalgebra
goes over to the ${\rm SU}(4)$ covariant form of the flat ${\cal N}\,{=}\,8, d\,{=}\,1$ superalgebra.
This limiting superalgebra actually possesses an enhanced $R$-symmetry group ${\rm SO}(8)$ which mixes $Q^I$ with $\bar{Q}_J$ (they are joined into ${\rm SO}(8)$ spinor).
In what follows we will not need the explicit form of these enhanced ${\rm SO}(8)/{\rm SU}(4)$ transformations, except for their realizations on the covariant
``flat' spinor derivatives.

The basic real ${\rm SU}(4|1)$\,, $d\,{=}\,1$ superspace is defined as the coset superspace
\bea
    \frac{{\rm SU}(4|1)}{{\rm SU}(4)}\sim \frac{\left\lbrace Q^{I}, \bar{Q}_{J}, L^I_J, {\cal H}\right\rbrace}{\left\lbrace L^I_J\right\rbrace}\,,\label{coset}
\eea
with the coset parameters being the superspace coordinates:
\bea
    \zeta = \left\lbrace t,\theta_{I},\bar{\theta}^{J}\right\rbrace ,\qquad
    \overline{\left(\theta_{I}\right)}=\bar{\theta}^{I}. \label{complexbasis_su41}
\eea
One could define these coordinates within the standard exponential parametrization of the supercoset.
However,  it will be more convenient to use another  parametrization, the one associated with the purely fermionic coset ${\rm SU}(n|1)/{\rm U}(n)$ defined in \cite{04cos} (see also \cite{04cos1}).
We uplift the ${\rm U}(1)$ group from the stability subgroup ${\rm U}(4)$ into the numerator and consider an extension of the ${\rm SU}(4|1)/{\rm U}(4)$ coordinate set by a time coordinate $t$.
Thus this ${\rm U}(1)$ generator is associated with the Hamiltonian.
Following to \cite{04cos}, one can then write generators of \eqref{algebra_su41} acting on the extended coset \eqref{coset} as
\bea
    &&Q^I=\frac{\partial}{\partial\theta_I}-2m\,\bar{\theta}^I\bar{\theta}^K\frac{\partial}{\partial\bar{\theta}^K}
    +i\bar{\theta}^I\partial_t\,,\qquad
    \bar{Q}_J=\frac{\partial}{\partial\bar{\theta}^J}
    +2m\,\theta_J\theta_K\,\frac{\partial}{\partial\theta_K}+i\theta_J \partial_t\,,\nn
    &&L^I_J=\left(\bar{\theta}^I\frac{\partial}{\partial\bar{\theta}^J}-\theta_J\,\frac{\partial}{\partial\theta_I}\right)
    -\frac{\delta^I_J}{4}\left(\bar{\theta}^K\frac{\partial}{\partial\bar{\theta}^K}-\theta_K\,\frac{\partial}{\partial\theta_K}\right),\nn
    &&{\cal H} = i\partial_{t}-\frac{3m}{4}\left(\bar{\theta}^K\frac{\partial}{\partial\bar{\theta}^K}
    -\theta_K\,\frac{\partial}{\partial\theta_K}\right). \label{diffgen}
\eea
Then, odd transformations corresponding to these supercharges are given by
\bea
    \delta\theta_{I} = \epsilon_{I} + 2m\,\bar{\epsilon}^{K}\theta_{K}\theta_{I}\,,\qquad
    \delta\bar{\theta}^{J} = \bar{\epsilon}^{J} - 2m\,\epsilon_{K}\bar{\theta}^{K}\bar{\theta}^{J},\qquad
    \delta t = i\left(\bar{\epsilon}^{K}\theta_{K} + \epsilon_{K}\bar{\theta}^{K}\right).\label{c_tr}
\eea
According to \cite{04cos}, one can define the integration measure as
\bea
    d\zeta :=   dt\,d^4\theta\,d^4\bar{\theta}\left(1 + 2 m\,\bar{\theta}^{K}\theta_{K}\right)^{3}. \label{inv}
\eea
It is easily checked to be invariant under the transformations \eqref{c_tr}.

Note that the Hamiltonian in \eqref{diffgen} is not a pure time derivative. One could pass to the new parametrization of superspace as
\bea
    \tilde{\theta}_{I} = \theta_{I}\,e^{3imt/4},\qquad
    \bar{\tilde{\theta}}^{I} = \bar{\theta}^{I}e^{-3imt/4},\qquad
    t=t\,,
\eea
in which the Hamiltonian takes the standard form ${\cal H} = i\partial_{t}$\,. The advantage of the parametrization \eqref{complexbasis_su41} is
the simplest form of the transformations \eqref{c_tr}. So, in what follows it will be convenient to deal with such a simple parametrization.
Due to the non-standard form of the Hamiltonian in this parametrization, all transformations and $\theta$-expansions of the ${\rm SU}(4|1)$
superfields will be accompanied by the factors like $e^{\pm 3imt/4}$.

\subsection{Chiral superspaces}
The supergroup ${\rm SU}(4|1)$ admits two mutually conjugated complex supercosets which can be identified with the left and right chiral subspaces:
\bea
    \zeta_{\rm L} = \left(t_{\rm L},\theta_I\right),\qquad  \zeta_{\rm R}= \left(t_{\rm R}, \bar{\theta}^J\,\right). \label{LR}
\eea
The left coordinate $t_{\rm L}$ is related to the real time coordinate $t$ via
\bea
    t_{\rm L} =t+\frac{i}{2m}\,\log{\left(1 + 2 m\,\bar{\theta}^K\theta_K\right)}\,.
\eea
Then we check that the left chiral space $\zeta_{\rm L}$ is closed under the supersymmetry transformations
\bea
    \delta\theta_{I}=\epsilon_{I}+2m\,\bar{\epsilon}^{K}\theta_{K}\theta_{I}\,,\qquad \delta t_{\rm L}=2i\bar{\epsilon}^K\theta_K\,.\label{left_tr}
\eea

The invariant left chiral measure is defined as
\bea
    d\zeta_{\rm L} := dt_{\rm L}\,d^4\theta\,e^{-3imt_{\rm L}},\qquad \delta\left(d\zeta_{\rm L}\right)=0\,, \qquad
&\int d\zeta_{\rm L}\,\theta_{I}\theta_{J}\theta_{K}\theta_{L}\,e^{3imt_{\rm L}} =\varepsilon_{IJKL}\,. \label{leftinv}
\eea

\subsection{Reduction to ${\rm SU}(2|1)$\,, $d\,{=}\,1$ superspace}
One can consider reduction of  the superspace \eqref{coset} to the ${\rm SU}(2|1)$ superspace.
It is performed on the superspace coordinates \eqref{complexbasis_su41} as
\bea
    \left\lbrace t,\theta_{i},\bar{\theta}^{i}\right\rbrace ,\qquad
    \overline{\left(\theta_{i}\right)}=\bar{\theta}^{i}, \quad i=1,2.\label{SU21_ss}
\eea
Limiting to the $\epsilon_1$ and $\epsilon_2$ transformations in \eqref{c_tr}, we obtain the reduced ${\rm SU}(2|1)$ supersymmetric transformations
which coincide with those found in \cite{DSQM}:
\bea
    \delta\theta_{i} = \epsilon_{i} + 2m\,\bar{\epsilon}^{k}\theta_{k}\theta_{i}\,,\quad
    \delta\bar{\theta}^{j} = \bar{\epsilon}^{j} - 2m\,\epsilon_{k}\bar{\theta}^{k}\bar{\theta}^{j},\quad
    \delta t = i\left(\bar{\epsilon}^{k}\theta_{k} + \epsilon_{k}\bar{\theta}^{k}\right).\label{SU21_tr}
\eea
Respectively, the superalgebra \eqref{algebra_su41} contains as a subalgebra the extended $su(2|1) +\!\!\!\!\!\!\supset u(1)$ superalgebra:
\bea
    &&\left\lbrace Q^{i}, \bar{Q}_{j}\right\rbrace = 2m\,I^{i}_{j} + m\,\delta^{i}_{j}F + 2\delta^{i}_{j}{\cal H},\qquad
    \left[I^i_j,  I^k_l\right] = \delta^k_j I^i_l - \delta^i_l I^k_j,\nn
    &&\left[I^i_j, Q^{k}\right]
    = \delta^k_k Q^{i} - \frac{1}{2}\,\delta^i_j Q^{k} ,\qquad \left[I^i_j, \bar{Q}_{l}\right] = \frac{1}{2}\,\delta^i_j\bar{Q}_{l}-\delta^i_l\bar{Q}_{j}\,,\nn
    &&\left[{\cal H}, Q^{k}\right]=-\,\frac{3m}{4}\,Q^{k},\qquad
    \left[{\cal H}, \bar{Q}_{l}\right]=\frac{3m}{4}\,\bar{Q}_{l}\,,\nn
    &&\left[F, Q^{k}\right]=\frac{1}{2}\,Q^{k},\qquad
    \left[F, \bar{Q}_{l}\right]=-\,\frac{1}{2}\,\bar{Q}_{l}\,.\label{algebra_su21}
\eea
Here, ${\rm SU}(2)$ generators of ${\rm SU}(2|1)$ are defined as
\bea
    I^i_j = L^i_j - \frac{1}{2}\,\delta^i_j F.
\eea
The combination ${\displaystyle{\cal H}+\frac{m}{2}\,F}$ can be identified with  the internal ${\rm U}(1)$ generator of ${\rm SU}(2|1)$,
while $F$ becomes an external $R$-symmetry ${\rm U}(1)$ generator.

The explicit expressions for the covariant spinor derivatives ${\cal D}^{k}, \bar{{\cal D}}^{k}$ corresponding to the basic real coset of ${\rm SU}(2|1)$ defined in \cite{ISTconf} and
parametrized by the coordinates \eqref{SU21_ss} with the transformation properties \eqref{SU21_tr} are given by
\bea
    {\cal D}^i &=& e^{-3imt/4}\bigg\lbrace\left[1+{m}\,\bar{\theta}^k\theta_k
    -\frac{3m^2}{8} \left(\theta\right)^2\left(\bar{\theta}\,\right)^2\right]\frac{\partial}{\partial\theta_i}
    - m\,\bar{\theta}^i\theta_j\frac{\partial}{\partial\theta_j}-i\bar{\theta}^i \partial_t\nn
    &&-\,\frac{m}{2}\,\bar{\theta}^i \tilde{F}- {m}\,\bar{\theta}^j\left(1 -m\,\bar{\theta}^k\theta_k \right)\tilde{I}^i_j\bigg\rbrace,\nn
    \bar{{\cal D}}_j &=& e^{3imt/4}\bigg\lbrace-\left[1+ {m}\,\bar{\theta}^k\theta_k
    -\frac{3m^2}{8} \left(\theta\right)^2\left(\bar{\theta}\,\right)^2\right]\frac{\partial}{\partial\bar{\theta}^j}
    + m\,\bar{\theta}^k\theta_j\frac{\partial}{\partial\bar{\theta}^k}+i\theta_j\partial_t\nn
    &&+\,\frac{m}{2}\,\theta_j\tilde{F}+ {m}\,\theta_k\left(1-m\,\bar{\theta}^l\theta_l \right)\tilde{I}^k_j\bigg\rbrace,\label{derivatives}
\eea
where
\bea
    &&\tilde{I}^i_j \bar{{\cal D}}_{l} = \delta^i_l\bar{{\cal D}}_{j}-\frac{1}{2}\,\delta^i_j\bar{{\cal D}}_{l}\, ,\qquad
    \tilde{I}^i_j {\cal D}^{k} =  \frac{1}{2}\,\delta^i_j {\cal D}^{k} -\delta^k_j {\cal D}^{i} ,\nn
    &&\tilde{F}\bar{{\cal D}}_{l}=\frac{1}{2}\,\bar{{\cal D}}_{l}\,,\qquad \tilde{F} {\cal D}^{k}=-\,\frac{1}{2}\,{\cal D}^{k}.
\eea
In what follows we will avoid using the explicit form of the ${\rm SU}(4|1)$ counterparts of these derivatives, though they can be
straightforwardly constructed by applying the standard coset (super)space machinery.

\section{${\rm SU}(4)$ covariant formulations of $(8, 8, 0)$ multiplet in flat ${\cal N}\,{=}\,8$ supersymmetry}
Prior to the discussion of the superfield description of the root ${\bf (8, 8, 0)}$ multiplets in ${\rm SU}(4|1)$ supersymmetry, we  will
consider ${\rm SU}(4)$ covariant form of its defining constraints in the standard flat ${\cal N}\,{=}\,8$ superspace,
bearing in mind that the deformation to ${\rm SU}(4|1)$ mechanics must respect $R$-symmetry ${\rm SU}(4)$\,.

 Such constraints can be written in the two superfield forms, both preserving not only ${\rm SU}(4)$ but also a non-manifest ${\rm SO}(8)$ $R$-symmetry.\footnote{In general,
 flat constraints defining the multiplet ${\bf (8,8,0)}$
 can be given many equivalent forms. For instance, in \cite{BIKL2}, they were written in ${\rm SU}(2)\times {\rm SU}(2)\times {\rm SU}(2)\times {\rm SU}(2)$ covariant form. The common feature of all these
 formulations is the hidden covariance of the constraints under the full $R$-symmetry group of ${\cal N}\,{=}\,8$ superalgebra, the group ${\rm SO}(8)$.}

In the first formulation one deals with a  chiral superfield $\Phi$ and an antisymmetric tensor superfield $Y^{IJ}$ satisfying the constraints
\footnote{For further  use, we introduce the antisymmetric tensor $\varepsilon^{IJKL}\equiv \varepsilon^{[IJKL]}\,$, such that
$$
    \varepsilon^{1234}=\varepsilon_{1234}=1\,,\qquad \varepsilon^{IJKL}\varepsilon_{IJKL}=24\,.
$$}
\bea
    && \bar{D}_J\,\Phi = 0\,, \qquad D^I\,\bar{\Phi} = 0\,, \qquad \bar{D}_I\bar{D}_J\,\bar{\Phi} = \frac{1}{2}\,\varepsilon_{IJKL}\,D^K D^L\,\Phi\,,\nn
    && \sqrt{2}\,D^I\,Y^{JK} = -\,\varepsilon^{IJKL}\,\bar{D}_L\,\bar{\Phi}\,,\qquad
    \sqrt{2}\,\bar{D}_J\,Y_{KL} = \varepsilon_{IJKL}\,D^I\,\Phi\,,\nn
    &&\overline{\left(Y^{IJ}\right)}=Y_{IJ}=\frac{1}{2}\,\varepsilon_{IJKL}\,Y^{KL},\qquad \overline{\left(\Phi\right)}=\bar{\Phi}\,,\label{constr_v1}
\eea
where the flat covariant derivatives are defined as
\bea
    D^I=\frac{\partial}{\partial\theta_I} - i\bar{\theta}^I\partial_t\,,\qquad
    \bar{D}_J=-\,\frac{\partial}{\partial\bar{\theta}^J}+i\theta_J \partial_t\,.
\eea
It is straightforward to check that \eqref{constr_v1} is covariant under the non-manifest ${\rm SO}(8)/{\rm SU}(4)$ symmetry transformations realized as
\bea
&&\delta D^I = -\,\sqrt{2}\,\Lambda^{IJ}\bar{D}_J+i\lambda\, D^I, \qquad \delta \bar{D}_J = \sqrt{2}\,\bar\Lambda_{IJ}D^I -i\lambda \,\bar{D}_J\,, \label{D_SO8U4} \\
&&\delta \Phi = -\,\bar\Lambda^{IJ}Y_{IJ}-2i\lambda\,\Phi\,, \qquad  \delta \bar{\Phi} = -\,\Lambda^{IJ}Y_{IJ}+2i\lambda\,\bar{\Phi}\,, \nn
&&\delta Y_{IJ} = \Lambda_{IJ}\,\Phi + \bar{\Lambda}_{IJ}\,\bar{\Phi}\,,
\eea
where the antisymmetric complex $4\times 4$ matrix
\bea
\Lambda_{IJ} = \frac12\,\varepsilon_{IJKL}\,\Lambda^{KL}, \qquad \bar\Lambda_{IJ} = \frac12\,\varepsilon_{IJKL}\,\bar{\Lambda}^{KL},
\eea
accommodates just $12$ real parameters of the coset ${\rm SO}(8)/{\rm U}(4)$ and $\lambda$ is the real ${\rm U}(1) \sim {\rm SO}(2)$ parameter.  One can check that indeed
\be
\Phi\bar\Phi + \frac12\,Y^{IJ} Y_{IJ}  = {\rm inv}\,.\label{qf1}
\ee

Another form of the ${\rm SU}(4)$ covariant superfield description of the multiplet ${\bf (8, 8, 0)}$ involves the general superfield $V^I$ which is subject to the constraints
\bea
    &&D^{I}V^{J} = \frac{1}{2}\,\varepsilon^{IJKL}\,\bar{D}_K\bar{V}_L\,,\qquad D^{(I}\,V^{J)} = 0\,,\qquad\bar{D}_{(K}\,\bar{V}_{L)}=0\,,\nn
    &&D^{I}\bar{V}_J  = \frac{1}{4}\,\delta^I_J D^{K}\bar{V}_K\,, \qquad \bar{D}_{J}V^{I}=\frac{1}{4}\,\delta^I_J\bar{D}_K V^{K}\qquad
    \overline{\left(V^I\right)}=\bar{V}_I\,.\label{constr_v2}
\eea
The non-manifest ${\rm SO}(8)/{\rm SU}(4)$ transformations of $V^I$ leaving covariant the system \eqref{constr_v2} are written this time as
\bea
     \delta V^I = \sqrt{2}\,\bar{\Lambda}^{IJ}\bar{V}_J-i\lambda\, V^I, \qquad
    \delta \bar{V}_J = -\,\sqrt{2}\,\Lambda_{IJ}V^I+i\lambda \,\bar{V}_J\,.
\eea
These transformations, together with the transformations of the covariant derivatives \eqref{D_SO8U4}, preserve
the constraints \eqref{constr_v2}. One can also see that
\bea
V^{I} \bar{V}_{I}  = {\rm inv}.\label{qf2}
\eea

It is rather easy to check that the constraints \eqref{constr_v1} leave in the bosonic sector of $\Phi, Y^{IJ}$ just the  complex bosonic field $\phi(t)$
and tensorial field $y^{IJ}(t)$ which are first
components of these superfields and transform as $\underline{\bf 1}$ and $\underline{\bf 6}$ of ${\rm SU}(4)$\,. The physical fermions are defined as $D^I\Phi|_{\theta = 0}$ and transform as
$\underline{\bf 4}$ of ${\rm SU}(4)$.  In the case of the constraints \eqref{qf2} the ${\rm SU}(4)$ assignment of the physical fields changes to the opposite:
the physical bosons are the first components of $V^J$ and transform as $\underline{\bf 4}$\,, while fermions are defined as $\bar{D}_K V^{K}|_{\theta=0}\,, D^K\bar{V}_K|_{\theta=0}\,, D^{[I} V^{J]}|_{\theta =0}$
and transform as $\underline{\bf 1} \oplus \underline{\bf 1}^{*} \oplus \underline{\bf 6}$\,. Thus, two ${\bf (8,8,0)}$ multiplets have ``inverted'' ${\rm SU}(4)$ contents:
the contents of bosons and fermions of the first version coincide with those of fermions and bosons in the second one.

In order to better understand the interplay between the two forms of the ${\bf (8, 8, 0)}$ multiplet, we note that the fermionic superfield $D^I\,\Phi$ transforms
precisely as $V^I$. It is easy to check that it satisfies the constraints \eqref{constr_v2} as a consequence of \eqref{constr_v1}. Analogously,
the fermionic superfields $-\,2\sqrt{2}\,D^IV^{J}$
and $D^{K}\bar{V}_K$ possess the same transformation properties as $Y^{IJ}$ and $\bar{\Phi}$\,,
respectively. It is also straightforward to check that such fermionic superfields satisfy \eqref{constr_v1} as a consequence of \eqref{constr_v2}.
In other words, by the first multiplet one can construct the ``derivative'' fermionic multiplet satisfying the Grassmann-odd version
of the second multiplet constraints  \eqref{constr_v2}.
After establishing this correspondence, we could consider \eqref{constr_v2} for some new independent Grassmann-even superfield $V^I$ and so come to the system \eqref{constr_v2}
as an alternative  description of the ${\bf (8, 8, 0)}$ multiplet with the same Grassmann parities for the component fields as in  the first version,
but with ``inverted'' ${\rm SU}(4)$ assignments  of these components. Its fermionic ``derivative''  satisfies the constraints \eqref{constr_v1}.

This interplay between two ${(\bf 8,8,0)}$ multiplets resembles a similar feature of  ``mirroring'' of ${\bf (4,4,0)}$ multiplets
in the standard  (flat) ${\cal N}\,{=}\,4$ mechanics \cite{DeldIv4}, \cite{FIS}. The bosonic and fermionic components of the  mutually mirror  ${\bf (4,4,0)}$ multiplets
form doublets with respect to different ${\rm SU}(2)$ factors of the full ${\rm SO}(4)$ $R$-symmetry group and are equivalent
up to switching the roles of these two commuting ${\rm SU}(2)$ groups. However, there is an essential difference. In the ${\cal N}\,{=}\,4$ case the bosonic fields of the
mutually mirror ${\bf (4,4,0)}$ multiplets are doublets of {\it different\/} $R$-symmetry ${\rm SU}(2)$ groups (the same is true for fermionic fields).
As is seen from \eqref{qf1}  and \eqref{qf2},
in the ${\cal N}\,{=}\,8$ case the relevant fields form 8-dimensional irreps of the {\it same\/} full $R$-symmetry ${\rm SO}(8)$
and  differ only in their assignments with respect to the fixed  ${\rm U}(4)\subset {\rm SO}(8)$. So these
two descriptions are associated with different embeddings of ${\rm U}(4)$ into ${\rm SO}(8)$.
The first version corresponds to splitting ${\rm SO}(8) \rightarrow {\rm SO}(2)\times {\rm SO}(6)$ and representing the SO(8)-multiplet of superfields
as a sum of ${\rm SO}(2)$ and ${\rm SO}(6)$ vectors.
Then ${\rm SU}(4)$ is identified with ${\rm SO}(6)$, the additional $R$-symmetry ${\rm U}(1)$ with
${\rm SO}(2)$, while  $\Phi$ and $Y^{IK}$ with the corresponding ${\rm SO}(2)$ and ${\rm SO}(6)$ vectors.
The second version corresponds to splitting ${\rm SO}(8) \rightarrow {\rm SO}(4)\times {\rm SO}(4)'$  and
representing the relevant SO(8) superfield set  as a sum of two 4-vectors. The diagonal ${\rm SO}(4)$ is identified with the ``minimally embedded'' ${\rm SO}(4) \subset {\rm SU}(4)$,
and two 4-vectors are joined into a complex fundamental spinor $V^I$ of ${\rm SU}(4)$.

Actually, the hidden ${\rm SO}(8)$ symmetry reveals the triality \cite{GSW1} between bosonic fields, fermionic fields and covariant derivatives. This triality interrelates the
three irreducible fundamental representations of ${\rm SO}(8)$, {\it viz.}  the vector representation and two spinorial ones.\footnote{To be more exact, the triality property
is inherent to the group Spin$(8)$.}
All three representations can be written in the ${\rm SU}(4)\times {\rm U}(1) \sim {\rm SO}(6)\times {\rm SO}(2)$ notation \cite{GSW2} as
\bea
    &&{\rm vector}\quad \underline{\bf 1}_{\,1} \oplus \underline{\bf 1}^{*}_{\,-1}  \oplus \underline{\bf 6}_{\,0}\,,\nn
    &&{\rm spinor}\quad \underline{\bf 4}_{\,1/2} \oplus \underline{\bf 4}^{*}_{\,-1/2}\,,\nn
    &&{\rm spinor}\quad \underline{\bf 4}_{\,-1/2} \oplus \underline{\bf 4}^{*}_{\,1/2}\,,
\eea
where the subscript index refers to the ${\rm U}(1)$ charge. Comparing this with the ${\rm U}(4)$ assignments of the bosonic and fermionic fields
of the $({\bf 8, 8, 0})$ multiplets, as well as of the covariant derivatives, we observe that just these ${\rm SO}(8)$ representations
are realized on the quantities in question.

Supposing that the roles of two spinor representations can be switched, in flat ${\cal N}\,{=}\,8$, $d\,{=}\,1$ supersymmetry
we can introduce yet a {\it third\/} multiplet ${\bf (8, 8, 0)}$
living on a different superspace,  with the covariant derivatives defined as
\bea
    \tilde{D}_{IJ}=\frac{1}{2}\,\varepsilon_{IJKL}\,\tilde{D}^{KL},\quad \overline{\left(\tilde{D}_{IJ}\right)}=\tilde{D}^{IJ},\qquad \tilde{D},\quad
    \bar{\tilde{D}}=\overline{\left(\tilde{D}\right)},
\eea
and so belonging to the vector representation. However, an ${\rm SU}(4)$ covariant formulation of this third ${\bf (8, 8, 0)}$ multiplet is beyond our purpose
because the  ${\rm SU}(4|1)$ covariant derivatives are ${\rm SU}(4)$ spinors by definition. So, this third option  does not admit a generalization to ${\rm SU}(4|1)$ supersymmetry,
in contrast to the first two.

In the case of the constraints \eqref{constr_v1}, the bosonic fields belong to the ${\rm SO}(8)$ vector representation, and the fermionic fields form ${\rm SO}(8)$ spinor.
For the multiplet given by \eqref{constr_v2} the picture is reversed, that is, the bosonic fields form an ${\rm SO}(8)$ spinor and the fermionic fields are combined into ${\rm SO}(8)$ vector.
So, from the standpoint of ${\rm SO}(8)$ $R$-symmetry, due to the  triality  property, both ${\bf (8, 8, 0)}$ multiplets can be considered as equivalent, once
the spinorial representation of the covariant spinor derivatives has been fixed and one deals with SO(8) invariant actions for these multiplets (for more detail, see Section~5.4).

The crucial point of the equivalence just discussed is the hidden ${\rm SO}(8)$ covariance of both sets of constraints \eqref{constr_v1} and \eqref{constr_v2}.
In the case of ${\rm SU}(4|1)$-deformed mechanics there is no ${\rm SO}(8)$ $R$-symmetry, only ${\rm U}(4)$ remains. For this reason one cannot expect the corresponding counterparts
of the two ``flat'' ${\bf (8, 8, 0)}$ multiplets to be equivalent to one another.\footnote{In the flat case the ${\cal N}\,{=}\,8$ supersymmetric Lagrangians are not obliged
to simultaneously respect the full ${\rm SO}(8)$ symmetry. So for ${\rm SO}(8)$ non-invariant Lagrangians  the equivalency of different (${\bf 8, 8, 0}$) multiplets may be broken
in the flat case too.}

\section{The ${\rm SU}(4|1)$ multiplet ${\bf (8,8,0)}$\,: first version}\label{v1}
The first version of the multiplet ${\bf (8,8,0)}$ is defined by the ${\rm SU}(4|1)$ covariant constraints
\bea
    && \bar{{\cal D}}_J\,\Phi = 0\,, \qquad {\cal D}^I\,\bar{\Phi} = 0\,, \qquad \bar{{\cal D}}_I\bar{{\cal D}}_J\,\bar{\Phi} = \frac{1}{2}\,\varepsilon_{IJKL}\,{\cal D}^K {\cal D}^L\,\Phi\,,\nn
    && \sqrt{2}\,{\cal D}^I\,Y^{JK} = -\,\varepsilon^{IJKL}\,\bar{{\cal D}}_L\,\bar{\Phi}\,,\qquad
    \sqrt{2}\,\bar{{\cal D}}_J\,Y_{KL} = \varepsilon_{IJKL}\,{\cal D}^I\,\Phi\,,\nn
    &&\overline{\left(Y^{IJ}\right)}=Y_{IJ}=\frac{1}{2}\,\varepsilon_{IJKL}\,Y^{KL},\qquad \overline{\left(\Phi\right)}=\bar{\Phi}\,,\label{880_1}
\eea
where $\Phi$ is a chiral superfield and $Y^{IJ}$ is an antisymmetric tensor superfield.
In the flat limit, when  $m\rightarrow 0$\,, $\bar{\cal D}_J \rightarrow \bar{D}_J$\,, ${\cal D}^I \rightarrow D^I$,
this set of constraints becomes the set of superfield constraints \eqref{constr_v1} defining the standard ${\cal N}\,{=}\,8$, $d\,{=}\,1$ multiplet ${\bf (8, 8, 0)}$ \cite{BIKL2},
such that only ${\rm SU}(4) \subset {\rm SO}(8)$ is manifest.

In what follows, we avoid calculation of the deformed covariant derivatives ${\cal D}^I$, $\bar{{\cal D}}_J$ (they in general involve complicated ${\rm U}(4)$ connection terms)
and consider the multiplet ${\bf (8,8,0)}$ in the chiral superspace description, harmonic superspace description and ${\rm SU}(2|1)$ superfield approach.

\subsection{Chiral superfield}
We consider the chiral superfield $\Phi$ given by the general $\theta$-expansion
\bea
    \Phi\left(t_{\rm L},\theta_{I}\right) &=& \phi + \sqrt{2}\,\theta_{K}\chi^{K}e^{3imt_{\rm L}/4} + \theta_{I}\theta_{J}A^{IJ}e^{3imt_{\rm L}/2} +
    \frac{\sqrt{2}}{3}\,\theta_{I}\theta_{J}\theta_{K}\xi^{IJK}e^{9imt_{\rm L}/4}\nn
    &&+\,\frac{1}{4}\,\varepsilon^{IJKL}\,\theta_{I}\theta_{J}\theta_{K}\theta_{L} B\, e^{3imt_{\rm L}},\quad
    A^{IJ}\equiv A^{[IJ]},\quad \xi^{IJK}\equiv \xi^{[IJK]}.\label{sf_ch}
\eea
The superfield $\Phi$ transforms as a singlet of the stability subgroup ${\rm SU}(4)$\,, {\it i.e.} $\delta_{su(4)} \Phi = 0$\,. Taking into account \eqref{left_tr}, we can
find the transformations of its components under the odd generators:
\bea
    && \delta \phi = -\,\sqrt{2}\,\epsilon_{K}\chi^{K}e^{3imt/4},\nn
    && \delta \chi^I = \sqrt{2}\,\bar{\epsilon}^I\left(i\dot{\phi}\right)e^{-3imt/4} - \sqrt{2}\,\epsilon_K A^{IK}e^{3imt/4},\nn
    && \delta A^{IJ} = 2\sqrt{2}\,\bar{\epsilon}^{[I}\left(i\dot{\chi}^{J]} + \frac{m}{4}\,\chi^{J]}\right)e^{-3imt/4} - \sqrt{2}\,\epsilon_{K}\xi^{IJK}e^{3imt/4},\nn
    && \frac{\sqrt{2}}{3}\,\delta\xi^{IJK} = 2\,\bar{\epsilon}^{[K}\left(i\dot{A}^{IJ]} + \frac{m}{2}\,A^{IJ]}\right)e^{-3imt/4} - \varepsilon^{IJKL}\,\epsilon_L B\, e^{3imt/4},\nn
    && \varepsilon^{IJKL}\,\delta B = \frac{8\sqrt{2}}{3}\,\bar{\epsilon}^{[L}\left(i\dot{\xi}^{IJK]} + \frac{3m}{4}\,\xi^{IJK]}\right)e^{-3imt/4}. \label{PreTran}
\eea
The general supersymmetric action can be written as a sum of integrals over chiral subspaces \cite{286}, \cite{su22} as
\bea
    S_{\rm chiral}= \int dt\,{\cal L}_{\rm chiral} = -\frac14 \left[\int d\zeta_{\rm L}\,K\left(\Phi\right)
    + \int d\zeta_{\rm R}\,\bar{K}\left(\bar{\Phi}\,\right)\right], \label{action_ch}
\eea
where the overall coefficient $-1/4$ is chosen for further convenience. The component form of this ${\rm SU}(4|1)$ invariant is found to be
\bea
    S_{\rm chiral} &=& -\frac14\,\int dt\,\bigg\lbrace 6B\, \partial_{\phi}K + \varepsilon_{IJKL}\left[\frac{2}{3}\,\chi^L\xi^{IJK}
     + \frac{1}{2}\,A^{IJ}A^{KL}\right]\left(\partial_{\phi}\right)^2 K\nn
    &&-\,\varepsilon_{IJKL}\,A^{IJ}\chi^K\chi^L \left(\partial_{\phi}\right)^3 K+\frac{1}{6}\,
    \varepsilon_{IJKL}\,\chi^I\chi^J\chi^K\chi^L \left(\partial_{\phi}\right)^4 K + \mbox{c.c.}\bigg\rbrace\,.\label{componentChir}
\eea

This invariant does not display the kinetic term of the fields in \eqref{sf_ch} and so  must be treated as a kind of ``pre-action'' for the ${\bf (8, 8, 0)}$ multiplet.
The genuine action appears after imposing some extra ${\rm SU}(4|1)$ covariant conditions on the components in \eqref{sf_ch}. Of course they should follow
from the rest of the superfield constraints \eqref{880_1}, but it is easier to guess their form directly at the component level,  requiring the final field content
to be ${\bf (8,8,0)}$ and resorting to the ${\rm SU}(4|1)$ covariance reasonings.

In this way we find that the components of the chiral superfield \eqref{sf_ch} must be  subjected to the following additional constraints
\bea
    &&A^{IJ} = \sqrt{2}\left(i\dot{y}^{IJ}-\frac{m}{2}\,y^{IJ}\right),\qquad \overline{\left(y^{IJ}\right)}=y_{IJ}=\frac{1}{2}\,\varepsilon_{IJKL}\,y^{KL},\nn
    &&\xi^{IJK} = -\,\varepsilon^{IJKL}\left(i\dot{\bar{\chi}}_L-\frac{5m}{4}\,\bar{\chi}_L\right),\qquad \overline{\left(\chi^I\right)}=\bar{\chi}_I\,,\nn
    &&B = \frac{2}{3}\left(\ddot{\bar{\phi}}+2im\dot{\bar{\phi}}\right).\label{constr880_2}
\eea
The odd ${\rm SU}(2|1)$ transformations are realized on this minimal set of fields as:
\bea
    &&\delta \phi = -\,\sqrt{2}\,\epsilon_{I}\chi^{I}e^{3imt/4},\qquad \delta \bar{\phi} = \sqrt{2}\,\bar{\epsilon}^{I}\bar{\chi}_{I}\,e^{-3imt/4},\nn
    &&\delta y^{IJ} = -\,2\,\bar{\epsilon}^{[I}\chi^{J]} e^{-3imt/4}+ \varepsilon^{IJKL}\epsilon_{K}\bar{\chi}_{L}\,e^{3imt/4},\nn
    &&\delta \chi^I = \sqrt{2}\,\bar{\epsilon}^I\left(i\dot{\phi}\right)e^{-3imt/4} - 2\,\epsilon_J\left(i\dot{y}^{IJ}-\frac{m}{2}\,y^{IJ}\right)e^{3imt/4},\nn
    &&\delta \bar{\chi}_I = -\,\sqrt{2}\,\epsilon_I\left(i\dot{\bar{\phi}}\right)e^{3imt/4} + 2\,\bar{\epsilon}^J\left(i\dot{y}_{IJ}+\frac{m}{2}\,y_{IJ}\right)e^{-3imt/4}. \label{tr880}
\eea
They are consistent with the transformations \eqref{PreTran} and leave invariant the constraints \eqref{constr880_2}.
\subsection{The final action}
Substituting the constraints \eqref{constr880_2} into the pre-action \eqref{componentChir}, we find the correct component Lagrangian in the form
\bea
    {\cal L}_{\rm SK} &=& g_1\left[\dot{\phi}\dot{\bar{\phi}}+\frac{1}{2}\,\dot{y}^{IJ}\dot{y}_{IJ}+ \frac{i}{2}\left(\chi^K\dot{\bar{\chi}}_K-\dot{\chi}^K\bar{\chi}_K\right)- \frac{5m}{4}\,\chi^K\bar{\chi}_K-\frac{m^2}{8}\,y^{IJ}y_{IJ}\right]\nn
    &&-\,\frac{im}{4}\left(\dot{\phi}\,\partial_{\phi}g_1 - \dot{\bar{\phi}}\,\partial_{\bar{\phi}}g_1\right)y^{IJ}y_{IJ}+2im\left(\dot{\phi}\,\partial_{\bar{\phi}}\bar{K}-\dot{\bar{\phi}}\,\partial_{\phi}K\right)\nn
    &&+\,\frac{1}{\sqrt{2}}\left(i\dot{y}_{IJ}-\frac{m}{2}\,y_{IJ}\right)\chi^I\chi^J\,\partial_{\phi}g_1+\frac{1}{\sqrt{2}}\left(i\dot{y}^{IJ}+\frac{m}{2}\,y^{IJ}\right)\bar{\chi}_I\bar{\chi}_J\, \partial_{\bar{\phi}}g_1\nn
    &&-\,\frac{i}{2}\left(\dot{\phi}\,\partial_{\phi}g_1 - \dot{\bar{\phi}}\,\partial_{\bar{\phi}}g_1\right)\chi^K\bar{\chi}_K - \frac{1}{24}\,\varepsilon^{IJKL}\,\bar{\chi}_I\bar{\chi}_J\bar{\chi}_K\bar{\chi}_L\,\partial_{\bar{\phi}}\partial_{\bar{\phi}}g_1\nn
    &&-\,\frac{1}{24}\,\varepsilon_{IJKL}\,\chi^I\chi^J\chi^K\chi^L\,\partial_{\phi}\partial_{\phi} g_1 \,.\label{L_SKM}
\eea
We observe that the complex fields $\phi$ parametrizes a special K\"ahler (SK) manifold with the metric
\bea
    g_1\left(\phi,\bar{\phi}\,\right) = \partial_{\phi} \partial_{\phi} K\left(\phi\right)+\partial_{\bar{\phi}}\partial_{\bar{\phi}}\bar{K}\left(\bar{\phi}\,\right).\label{SKM}
\eea

\subsection{Supercharges}
The matrix models based on the multiplet under consideration, in the case of the simplest target space
metric $g=1$ (i.e for the free  model), were studied in \cite{Motl}. Here, we consider
a one-particle model generalized to the general SK metric
\eqref{SKM} and find the relevant classical ${\rm SU}(4|1)$ supercharges. Poisson (Dirac)
brackets are written as
\bea
    &&\left\lbrace \phi, p_{\phi}\right\rbrace = 1,\qquad\left\lbrace \bar{\phi}, p_{\bar\phi}\right\rbrace
    = 1,\qquad \left\lbrace y^{KL}, p_{IJ}\right\rbrace = \frac{1}{2}\left(\delta^K_I\delta^L_J-\delta^L_I\delta^K_J\right),\nn
    &&\left\lbrace \chi^I ,\bar{\chi}_J\right\rbrace = -\,i\,\delta^I_J\left(g_1\right)^{-1},\nn
    &&\left\lbrace p_{\phi},\chi^I\right\rbrace = \frac{1}{2}\left(g_1\right)^{-1}\partial_{\phi}g_1\,\chi^I,\qquad
    \left\lbrace p_{\bar\phi},\chi^I\right\rbrace = \frac{1}{2}\left(g_1\right)^{-1}\partial_{\bar\phi}g_1\,\chi^I,\nn
    &&\left\lbrace p_{\phi},\bar{\chi}_J\right\rbrace = \frac{1}{2}\left(g_1\right)^{-1}\partial_{\phi}g_1\,\bar{\chi}_J\,,\qquad
    \left\lbrace p_{\bar\phi},\bar{\chi}_J\right\rbrace = \frac{1}{2}\left(g_1\right)^{-1}\partial_{\bar\phi}g_1\,\bar{\chi}_J\,.
    \label{brackets}
\eea
Then the Noether supercharges are given by
\bea
    Q^I &=& e^{3imt/4}\,\bigg\lbrace\,2\bar{\chi}_K\left[p^{IK}+\frac{i}{2}\,m\,g_1\,y^{IK}-\frac{i}{6\sqrt{2}}\,
    \varepsilon^{IKLM}\bar{\chi}_L\bar{\chi}_M\, \partial_{\bar{\phi}}g_1\right]\nn
    &&-\,\sqrt{2}\,\chi^I\left(p_{\phi}-2im\,\partial_{\bar\phi}\bar{K}+\frac{i}{4}\,m\,\partial_{\phi}g_1\,y^{KL}y_{KL}
    +\frac{i}{2}\,\partial_{\phi}g_1\,\chi^K\bar{\chi}_K\right)\bigg\rbrace\,,\nn
    \bar{Q}_J &=& e^{-3imt/4}\,\bigg\lbrace\,2\chi^K\left[p_{JK}-\frac{i}{2}\,m\,g_1\,y_{JK}-\frac{i}{6\sqrt{2}}\,
    \varepsilon_{JKLM}\chi^L\chi^M\,\partial_{\phi}g_1\right]\nn
    &&-\,\sqrt{2}\,\bar{\chi}_J\left(p_{\bar\phi}+2im\,\partial_{\phi}K-\frac{i}{4}\,m\,\partial_{\bar\phi}g_1\,y^{KL}y_{KL}
    -\frac{i}{2}\,\partial_{\bar{\phi}}g_1\,\chi^K\bar{\chi}_K\right)\bigg\rbrace\,.\label{supercharges}
\eea
Taking into account the brackets \eqref{brackets}, these
supercharges close on the following bosonic generators
\bea
    {\cal H}_{\rm SK} &=&\left(g_1\right)^{-1}\left(p_{\phi}-2im\,\partial_{\bar\phi}\bar{K}
    +\frac{i}{4}\,m\,\partial_{\phi}g_1\,y^{IJ}y_{IJ}+\frac{i}{2}\,\partial_{\phi}g_1\,\chi^K\bar{\chi}_K\right)\nn
    &&\times\left(p_{\bar\phi}+2im\,\partial_{\phi}K-\frac{i}{4}\,m\,\partial_{\bar\phi}g_1\,y^{IJ}y_{IJ}-\frac{i}{2}\,\partial_{\bar\phi}g_1\,\chi^K\bar{\chi}_K\right)\nn
    &&+\,\frac{1}{2g_1}\left[p^{IJ}-\frac{i}{\sqrt{2}}\left(\chi^I\chi^J\partial_{\phi}g_1+\frac{1}{2}\,\varepsilon^{IJKL}\bar{\chi}_K\bar{\chi}_L\,\partial_{\bar\phi}g_1\right)\right]\nn
    &&\times\left[p_{IJ}-\frac{i}{\sqrt{2}}\left(\bar{\chi}_I\bar{\chi}_J\,\partial_{\bar\phi}g_1+\frac{1}{2}\,\varepsilon_{IJMN}\,\chi^M\chi^N\partial_{\phi}g_1\right)\right]\nn
    &&+\,g_1\left[\frac{5m}{4}\,\chi^K\bar{\chi}_K+\frac{m^2}{8}\,y^{IJ}y_{IJ}\right]
    +\frac{m}{2\sqrt{2}}\left(y_{IJ}\chi^I\chi^J\,\partial_{\phi}g_1-y^{IJ}\bar{\chi}_I\bar{\chi}_J\, \partial_{\bar{\phi}}g_1\right)\nn
    &&+\,\frac{1}{24}\,\varepsilon^{IJKL}\,\bar{\chi}_I\bar{\chi}_J\bar{\chi}_K\bar{\chi}_L\,\partial_{\bar{\phi}}\partial_{\bar{\phi}}g_1
    +\frac{1}{24}\,\varepsilon_{IJKL}\,\chi^I\chi^J\chi^K\chi^L\,\partial_{\phi}\partial_{\phi} g_1 \,,\\
    L^I_J &=& 2i\left(y^{IK}p_{JK} - \frac{\delta^I_J}{4}\,y^{KL}p_{KL}\right)+
    g_1\left(\chi^I\bar{\chi}_J - \frac{\delta^I_J}{4}\,\chi^K\bar{\chi}_K\right),
\eea
in full agreement with the superalgebra \eqref{algebra_su41}. The quantum version of these ${\rm SU}(4|1)$ (super)charges can be straightforwardly constructed and will be presented elsewhere.

\subsection{Harmonic superspace description}
We consider the harmonic coset of ${\rm SU}(4|1)$ with the harmonic part  $\frac{{\rm SU}(4)}{[{\rm SU}(2)\times {\rm SU}(2)\times {\rm U}(1)]}$ \cite{nguyen}. The relevant harmonic variables
are $u^{(+)I}_{a}$, $u^{(+)i}_I$, $u^{(-)a}_{I}$, $u^{(-)I}_{i}$ where $i=1,2$ and $a=1,2$ are the indices of the fundamental representations of the subgroup ${\rm SU}(2)\times {\rm SU}(2)$.
The unitarity and unimodularity conditions are written as
\bea
    &&  u^{(+)i}_{K}u^{(-)K}_{j} = \delta^{i}_{j}\,,\quad
    u^{(-)a}_{K} u^{(+)K}_{b} = \delta^{a}_{b}\,,\qquad  u^{(-)a}_{J} u^{(+)I}_{a} +u^{(+)i}_{J}u^{(-)I}_i = \delta^{I}_{J}\,,\nn
    &&  u^{(-)a}_{K}u^{(-)K}_{j} = u^{(+)i}_{K} u^{(+)K}_{b} = 0\,,\quad
    \varepsilon^{IJKL}\varepsilon_{ij}\,u^{(+)i}_{K}u^{(+)j}_{L} + 2\,\varepsilon^{ab}u^{(+)I}_{a}u^{(+)J}_{b}=0\,.
\eea
Defining the harmonic projections of the ${\rm SU}(4|1)$ Grassmann coordinates as
\bea
    &&\theta^{(+)}_{a} = \theta_{I}\left(u^{(+)I}_{a}+m\,\bar{\theta}^{(+)k}\theta^{(+)}_{a}u^{(-)I}_{k}\right) ,\qquad \theta^{(-)}_{i} = \theta_{I}\,u^{(-)I}_{i},\nn
    &&\bar{\theta}^{(+)i} = \bar{\theta}^{J}\left(u^{(+)i}_{J}+m\,\bar{\theta}^{(+)i}\theta^{(+)}_{c}u^{(-)c}_{J}\right), \qquad \bar{\theta}^{(-)a} = \bar{\theta}^{J}u^{(-)a}_J\,,\nn
    &&t_{\rm A} = t + i\left(\bar{\theta}^{(-)a} \theta^{(+)}_{a} - \bar{\theta}^{(+)i}\theta^{(-)}_i\right)\left[1 - m\,\left(\bar{\theta}^{(-)a} \theta^{(+)}_{a} + \bar{\theta}^{(+)i}\theta^{(-)}_i\right)\right],
\eea
one can find that they transform  as
\bea
    &&\delta \theta^{(-)}_{i} = \epsilon^{(-)}_{i} + 2m\left[\bar{\epsilon}^{(-)c}\theta^{(+)}_{c}\left(1+m\,\bar{\theta}^{(+)k}\theta^{(-)}_{k}\right)+\bar{\epsilon}^{(+)k}\theta^{(-)}_{k}\right]\theta^{(-)}_{i},\nn
    &&\delta \bar{\theta}^{(-)a} = \bar{\epsilon}^{(-)a} - 2m\left[\epsilon^{(+)}_{c}\bar{\theta}^{(-)c}+\epsilon^{(-)}_{k}\bar{\theta}^{(+)k}\left(1+m\,\bar{\theta}^{(-)c}\theta^{(+)}_{c}\right)\right]\bar{\theta}^{(-)a},\nn
    &&\delta \theta^{(+)}_{a} = \epsilon^{(+)}_{a}+m\,\epsilon^{(-)}_{k}\bar{\theta}^{(+)k}\theta^{(+)}_{a}+2m\,\bar{\epsilon}^{(-)c}\theta^{(+)}_{c}\theta^{(+)}_{a},\nn
    &&\delta \bar{\theta}^{(+)i} = \bar{\epsilon}^{(+)i}-m\,\bar{\epsilon}^{(-)c}\theta^{(+)}_{c}\bar{\theta}^{(+)i}-2m\,\epsilon^{(-)}_{k}\bar{\theta}^{(+)k}\bar{\theta}^{(+)i},\nn
    &&\delta u^{(+)i}_{I} = -\,\Lambda^{(+2)i}_{b}u^{(-)b}_{I},\qquad \delta u^{(-)I}_{i} = 0\,,\nn
    &&\delta u^{(+)I}_{b} = \Lambda^{(+2)i}_{b} u^{(-)I}_{i},\qquad \delta u^{(-)b}_{I} = 0\,,\nn
    &&\delta t_{\rm A} = 2i\left(\epsilon^{(-)}_{k}\bar{\theta}^{(+)k}+\bar{\epsilon}^{(-)c}\theta^{(+)}_{c}\right),\label{HSSII_tr}
\eea
where
\bea
    &&\Lambda^{(+2)i}_{a} = m\left(\epsilon^{(+)}_{a}\bar{\theta}^{(+)i}+\bar{\epsilon}^{(+)i}\theta^{(+)}_{a}\right)+m^2\left(\epsilon^{(-)}_{k}\bar{\theta}^{(+)k}+\bar{\epsilon}^{(-)c}\theta^{(+)}_{c}\right)\bar{\theta}^{(+)i}\theta^{(+)}_{a},\nn
    &&\epsilon^{(-)}_{i} = \epsilon_{I}\,u^{(-)I}_{i} , \qquad \epsilon^{(+)}_{a} = \epsilon_{I}\,u^{(+)I}_{a},\qquad
    \bar{\epsilon}^{(+)i} = \bar{\epsilon}^{J}u^{(+)i}_{J}, \qquad \bar{\epsilon}^{(-)a} = \bar{\epsilon}^{J}u^{(-)a}_J.\qquad
\eea
We observe the existence of the analytic subspace closed under the ${\rm SU}(4|1)$ supersymmetry
\bea
    \zeta_{\rm A} =\left\lbrace t_{\rm A}, \theta^{(+)}_{a}, \bar{\theta}^{(+)i}, u^{(+)I}_{a}, u^{(+)i}_I, u^{(-)a}_{I}, u^{(-)I}_{i}\right\rbrace .\label{HSSII}
\eea
Its integration measure is given by
\bea
    &&d\zeta_{\rm A}^{(-4)} = dt_{\rm A}\,du\,d^2\theta^{(+)}\,d^2\bar{\theta}^{(+)}\quad \Rightarrow \;\nn
    && \; \Rightarrow \quad\delta\left(d\zeta_{\rm A}^{(-4)}\right) = 2\,d\zeta_{\rm A}^{(-4)}\Lambda^{(0)},\qquad
    \Lambda^{(0)} = m\left(\epsilon^{(-)}_{k}\bar{\theta}^{(+)k}-\bar{\epsilon}^{(-)c}\theta^{(+)}_{c}\right).\label{measure2}
\eea
The only harmonic derivative ${\cal D}^{(+2)i}_{a}$ preserving the analytic subspace reads
\bea
    {\cal D}^{(+2)i}_{a} &=& u^{(+)K}_{a}\,\frac{\partial}{\partial u^{(-)K}_{i}} - u^{(+)i}_{K}\,\frac{\partial}{\partial u^{(-)a}_{K}} - 2i\bar{\theta}^{(+)i}\theta^{(+)}_{a}\partial_{\rm A}\nn
    &&+\,m\,\bar{\theta}^{(+)i}\theta^{(+)}_{a}\left( \bar{\theta}^{(+)k}\,\frac{\partial}{\partial \bar{\theta}^{(+)k}}-\theta^{(+)}_{c}\,\frac{\partial}{\partial \theta^{(+)}_{c}} \right)\nn
    &&+\,\frac{m^2}{4}\,\varepsilon^{ij}\varepsilon_{ab}\left(\theta^{(+)}\right)^4\left(u^{(-)K}_{j}\,\frac{\partial}{\partial u^{(+)K}_{b}} - u^{(-)b}_{K}\,\frac{\partial}{\partial u^{(+)j}_{K}}\right),
\eea
where
\bea
    \left(\theta^{(+)}\right)^4=\left(\bar{\theta}^{(+)}\right)^2 \left(\theta^{(+)}\right)^2 = \bar{\theta}^{(+)k}\bar{\theta}^{(+)}_{k}\theta^{(+)}_{c}\theta^{(+)c}.
\eea
The remaining harmonic covariant derivatives prove undeformed:
\bea
    {\cal D}^{(0)} &=& u^{(+)k}_{K}\,\frac{\partial}{\partial u^{(+)k}_{K}} + u^{(+)K}_{c}\,\frac{\partial}{\partial u^{(+)K}_{c}} -  u^{(-)K}_{k}\,\frac{\partial}{\partial u^{(-)K}_{k}} - u^{(-)c}_{K}\,\frac{\partial}{\partial u^{(-)c}_{K}}\nn
    && +\,\theta^{(+)}_{c}\,\frac{\partial}{\partial \theta^{(+)}_{c}} + \bar{\theta}^{(+)k}\,\frac{\partial}{\partial \bar{\theta}^{(+)k}}\,,\nn
    {\cal D}^{i}_{j} &=& u^{(+)i}_{K}\,\frac{\partial}{\partial u^{(+)j}_{K}} - u^{(-)K}_{j}\,\frac{\partial}{\partial u^{(-)K}_{i}} + \bar{\theta}^{(+)i}\,\frac{\partial}{\partial \bar{\theta}^{(+)j}}\nn
    &&-\,\frac{\delta^i_j}{2}\left(u^{(+)k}_{K}\,\frac{\partial}{\partial u^{(+)k}_{K}} - u^{(-)K}_{k}\,\frac{\partial}{\partial u^{(-)K}_{k}}+\bar{\theta}^{(+)k}\,\frac{\partial}{\partial \bar{\theta}^{(+)k}}\right),\nn
    {\cal D}^{a}_{b} &=& u^{(-)a}_{K}\,\frac{\partial}{\partial u^{(-)b}_{K}}-u^{(+)K}_{b}\,\frac{\partial}{\partial u^{(+)K}_{a}} - \theta^{(+)}_{b}\,\frac{\partial}{\partial \theta^{(+)}_{a}}\nn
    &&-\,\frac{\delta^{a}_{b}}{2}\left(u^{(-)c}_{K}\,\frac{\partial}{\partial u^{(-)c}_{K}}-u^{(+)K}_{c}\,\frac{\partial}{\partial u^{(+)K}_{c}}-\theta^{(+)}_{c}\,\frac{\partial}{\partial \theta^{(+)}_{c}}\right),
\eea
One can check that
\bea
    &&\Lambda^{(+2)i}_{a} = {\cal D}^{(+2)i}_{a}\Lambda^{(0)},\qquad \varepsilon^{ab}\varepsilon_{ij}\,{\cal D}^{(+2)j}_{b}\Lambda^{(+2)i}_{a}= m^2\,\delta\left(\theta^{(+)}\right)^4,\nn
    &&\delta\left(\theta^{(+)}\right)^4 = 2\left(\delta\bar{\theta}^{(+)k}\right)\bar{\theta}^{(+)}_{k}\theta^{(+)}_{c}\theta^{(+)c}+2\,\bar{\theta}^{(+)k}\bar{\theta}^{(+)}_{k}\theta^{(+)}_{c}\left(\delta\theta^{(+)c}\right).
\eea
and
\bea
    \delta {\cal D}^{(+2)i}_{a} = \Lambda^{(+2)i}_{c}{\cal D}^{c}_{a}-\Lambda^{(+2)k}_{a}{\cal D}^{i}_{k}-\frac{\Lambda^{(+2)i}_{a}}{2}\,{\cal D}^{(0)}.
\eea
\subsubsection{Analytic harmonic superfield}
The relevant analytic harmonic superfield is defined by the conditions
\bea
    {\cal D}^{(+2)i}_{a}Y^{(+2)}=0\,,\qquad {\cal D}^{i}_{j}Y^{(+2)}={\cal D}^{a}_{b}Y^{(+2)}=0\,,\label{Y+2_constr}
\eea
and it transforms as
\bea
    \delta Y^{(+2)}=\Lambda^{(0)}Y^{(+2)}.
\eea
It can be obtained by the ``harmonization'' of the superfield $Y^{IJ}$ satisfying the constraints
\bea
    {\cal D}^{(K}Y^{I)J}=0\,,\qquad \bar{\cal D}_{(K}\,Y_{I)J}=0\,.
\eea
These constraints in fact define the multiplet ${\bf (6,8,2)}$. On the other hand, they are part of the full set of the constraints \eqref{880_1}
defining the multiplet ${\bf (8,8,0)}$\,.
Indeed, the solution of \eqref{Y+2_constr}
\bea
    Y^{(+2)} &=& y^{(+2)}+2i\bar{\theta}^{(+)i}\theta^{(+)a}\dot{y}_{ia}-\bar{\theta}^{(+)k}\bar{\theta}^{(+)}_{k}\theta^{(+)}_{a}\theta^{(+)a}\ddot{y}^{(-2)}\nn
    &&+ \,\bar{\theta}^{(+)}_iu^{(+)i}_{K}\chi^K e^{-3imt_{\rm A}/4}+\bar{\theta}^{(+)i}\bar{\theta}^{(+)}_{i}\theta^{(+)}_{a}u^{(-)a}_{K}\left(i\dot{\chi}^K+\frac{m}{4}\,\chi^K\right)e^{-3imt_{\rm A}/4}\nn
    &&  + \,\theta^{(+)a}u^{(+)K}_{a}\bar{\chi}_K\,e^{3imt_{\rm A}/4} +\theta^{(+)}_{a}\theta^{(+)a}\bar{\theta}^{(+)i}u^{(-)K}_{i}\left(i\dot{\bar{\chi}}_K-\frac{m}{4}\,\bar{\chi}_K\right)e^{3imt_{\rm A}/4}\nn
        && +\,\frac{1}{\sqrt{2}}\left(\bar{\theta}^{(+)i}\bar{\theta}^{(+)}_{i}D\,e^{-3imt_{\rm A}/2} +\theta^{(+)}_{a}\theta^{(+)a}\bar{D}\,e^{3imt_{\rm A}/2}\right),
\eea
reveals the field content ${\bf (6,8,2)}$,  where
\bea
    &&y^{(+2)}=\frac{1}{2}\,\varepsilon^{ab}\,u^{(+)I}_{a}u^{(+)J}_{b}y_{IJ}+\frac{m^2}{4}\left(\theta^{(+)}\right)^4y^{(-2)}\,,\nn
    &&y_{ia}=u^{(+)I}_{a}u^{(-)J}_{i}y_{IJ}\,,\quad
    y^{(-2)}=\frac{1}{2}\,\varepsilon^{ij}\,u^{(-)I}_{i}u^{(-)J}_{j}y_{IJ}\,.
\eea
The component fields transform as
\bea
    &&\delta D = -\,\sqrt{2}\,\epsilon_{I}\left(i\dot{\chi}^{I}-\frac{3m}{4}\,\chi^{I}\right)e^{3imt/4},\qquad \delta \bar{D} = -\,\sqrt{2}\,\bar{\epsilon}^{I}\left(i\dot{\bar{\chi}}_{I}+\frac{3m}{4}\,\bar{\chi}_{I}\right)e^{-3imt/4},\nn
    &&\delta y^{IJ} = -\,2\,\bar{\epsilon}^{[I}\chi^{J]} e^{-3imt/4}+ \varepsilon^{IJKL}\epsilon_{K}\bar{\chi}_{L}\,e^{3imt/4},\nn
    &&\delta \chi^I = \sqrt{2}\,\bar{\epsilon}^I D\, e^{-3imt/4} - 2\,\epsilon_J\left(i\dot{y}^{IJ}-\frac{m}{2}\,y^{IJ}\right)e^{3imt/4},\nn
    &&\delta \bar{\chi}_I = \sqrt{2}\,\epsilon_I \bar{D}\,e^{3imt/4} + 2\,\bar{\epsilon}^J\left(i\dot{y}_{IJ}+\frac{m}{2}\,y_{IJ}\right)e^{-3imt/4}. \label{tr682}
\eea
The substitution $D=i\dot{\phi}$ in these transformations gives just the transformations \eqref{tr880} of the multiplet ${\bf (8,8,0)}$\,.
Thus this substitution  ensures the validity of the additional constraints imposed on the superfield $Y^{IJ}$. We conclude that the ${\rm SU}(4|1)$ multiplet ${\bf (8,8,0)}$ admits an alternative description
within harmonic ${\rm SU}(4|1)$ superspace.

\subsubsection{Invariant action via harmonic superspace}\label{action}
Introducing the shifted superfield
\bea
    Y^{(+2)}&=&\hat{Y}^{(+2)}+c^{(+2)},\nn
    c^{(+2)}&=&\frac{1}{2}\,\varepsilon^{ab}\,u^{(+)I}_{a}u^{(+)J}_{b}c_{IJ}+\frac{m^2}{4}\left(\theta^{(+)}\right)^4c^{(-2)},\qquad c^{(-2)}=\frac{1}{2}\,\varepsilon^{ij}\,u^{(-)I}_{i}u^{(-)J}_{j}c_{IJ}\,,\nn
    \delta \hat{Y}^{(+2)}&=&\Lambda^{(0)}\left(\hat{Y}^{(+2)}+c^{(+2)}\right)+\varepsilon^{ab}\varepsilon_{ij}\,\Lambda^{(+2)i}_{a}\left({\cal D}^{(+2)j}_{b}c^{(-2)}\right)\nn
    &&-\,\frac{1}{4}\,\varepsilon^{ab}\varepsilon_{ij}\,c^{(-2)}\left({\cal D}^{(+2)j}_{b}\Lambda^{(+2)i}_{a}\right),\label{hatY}
\eea
we calculate the invariant action (see Appendix \ref{calc}) as
\bea
    S_{\bf (6,8,2)} &=& \frac{1}{16}\,\int\,d\zeta_{\rm A}^{(-4)}L^{(+4)},\nn
    L^{(+4)} &=& \frac{\hat{Y}^{(+2)}\hat{Y}^{(+2)}}{\left(1+c^{(-2)}\hat{Y}^{(+2)}\right)^4}
    +\frac{m^2}{2}\left(\theta^{(+)}\right)^4\left[1-\frac{1-c^{(-2)}\hat{Y}^{(+2)}}{\left(1+c^{(-2)}\hat{Y}^{(+2)}\right)^5}\right].\label{L+4act}
\eea
This action of the multiplet ${\bf (6,8,2)}$ is in fact superconformal with respect to the supergroup ${\rm SU}(4|1,1)$ (see Appendix \ref{682}).
The relevant metric is ${\rm SO}(6)$ invariant and given by
\bea
    g_2 = \left[\frac{1}{2}\,y^{IJ}y_{IJ}\right]^{-2}.\label{g_so6}
\eea
Substituting $D=i\dot{\phi}$, one can finally find the bosonic truncation of the component Lagrangian for the multiplet ${\bf (8,8,0)}$:
\bea
    {\cal L}_{\rm bos.} &=& g_2\left(\dot{\phi}\dot{\bar{\phi}}+\frac{1}{2}\,\dot{y}^{IJ}\dot{y}_{IJ}-\frac{m^2}{8}\,y^{IJ}y_{IJ}\right).
\eea
Calculation of all terms in harmonic superspace is rather complicated. We skip all these calculations and write
the full component Lagrangian \eqref{so6} in the next subsection by employing ${\rm SU}(2|1)$ superfields.

\subsection{${\rm SU}(2|1)$ superfield approach}

To simplify the construction of ${\rm SU}(4|1)$ invariant actions, it  will be convenient to employ ${\rm SU}(2|1)$ superfield approach elaborated in \cite{DSQM}, \cite{SKO}, \cite{ISTconf}, \cite{DHSS}.
We split the multiplet ${\bf (8,8,0)}$ into ${\rm SU}(2|1)$ multiplets as a sum of the conventional multiplet ${\bf (4,4,0)}$ and the ``mirror'' multiplet ${\bf (4,4,0)}$ \cite{DHSS}.
To obtain such a decomposition, we need to single out the $\epsilon_1$ and $\epsilon_2$ subvariety of the transformations of \eqref{tr880} corresponding
to the ${\rm SU}(2|1)$ superspace transformations \eqref{SU21_tr}. The ${\rm SU}(2|1)$ covariant constraints given below involve the covariant derivatives \eqref{derivatives}.

\subsubsection{The standard multiplet ${\bf (4,4,0)}$}
Introducing the new notations
\bea
    &&x^{11}:=y^{14},\quad x^{12}:=y^{13},\quad x^{21}:=y^{24},\quad x^{22}:=y^{23},\nn
    &&\xi^1 := \bar{\chi}_3\,,\quad \xi^2 := -\,\bar{\chi}_4\,,\qquad \bar{\xi}_1 := \chi^3,\quad\bar{\xi}_2 := -\,\chi^4 ,\nn
    &&\overline{\left(x^{ia}\right)} = \varepsilon_{ab}\,\varepsilon_{ij}\,x^{jb},
\qquad
    \overline{\left(\xi^a\right)} = \bar{\xi}_a\,,\qquad
    \overline{\left(\xi^{a}\right)} = \bar{\xi}_{a}\,,
\eea
we obtain the same deformed transformations as in \cite{DHSS}:
\bea
    &&\delta x^{ia}=-\left(\epsilon^{i}\xi^{a}e^{3imt/4}+\bar{\epsilon}^{i}\bar{\xi}^{a}e^{-3imt/4}\right),\nn
    &&\delta \xi^a = \bar{\epsilon}^k \left(2i\dot{x}_{k}^{a}+m\,x_{k}^{a}\right)e^{-3imt/4},\qquad
    \delta \bar{\xi}^a = \epsilon_k \left(2i\dot{x}^{ka}-m\,x^{ka}\right)e^{3imt/4}.
\eea
The indices $i\,{=}\,1,2$ and $a\,{=}\,1,2$ correspond to the fundamental representations of the subgroup ${\rm SU}(2)\times {\rm SU}(2)\subset {\rm SU}(4)$\,.

The corresponding superfield $q^{ia}$ obeys the ${\rm SU}(2|1)$ covariant constraints
\bea
    {\cal D}^{(k}q^{i)a} = \bar{{\cal D}}^{(k}q^{i)a} =0\,, \qquad \tilde{F}q^{ia} = 0\,, \qquad \overline{\left(q^{ia}\right)} = q_{ia}\,.\label{440constr1}
\eea
These constraints are solved by
\bea
    q^{ia}&=&\left[1+ \frac{m}{2}\,\bar{\theta}^k\theta_k - \frac{5m^2}{16}\left(\bar{\theta}\,\right)^2\left(\theta\right)^2\right]x^{ia}
    +\left(1+\frac{m}{4}\,\bar{\theta}^k\theta_k\right)\left(\theta^i \xi^a e^{3imt/4}
    +\bar{\theta}^i\bar{\xi}^a e^{-3imt/4}\right)\nn
    && +\,i\left( \bar{\theta}^k \theta^i\dot{x}^{a}_{k}-\bar{\theta}^i\theta_k\dot{x}^{ka}\right)
    -i\bar{\theta}^k\theta_k\left(\theta^i \dot{\xi}^a e^{3imt/4}
    -\bar{\theta}^i\dot{\bar{\xi}}^a e^{-3imt/4}\right) +\frac{1}{4}\left(\bar{\theta}\,\right)^2\left(\theta\right)^2\ddot{x}^{ia},\nn
    \label{q}
\eea
where the following conventions for the Grassmann monomials were employed: $\left(\theta\right)^2 = \theta_i\theta^i$, $\left(\bar{\theta}\,\right)^2 = \bar{\theta}^i\bar{\theta}_i$\,.

\subsubsection{The mirror multiplet ${\bf (4,4,0)}$}
The ``mirror'' ${\bf (4,4,0)}$ multiplet is defined by the transformations
\bea
    && \delta z = -\,\epsilon_{k}\psi^{k}e^{3imt/4},\quad \delta \bar{z} = \bar{\epsilon}^{k}\bar{\psi}_{k}\,e^{-3imt/4},\nn
    &&\delta y = -\,\epsilon_{k}\bar{\psi}^{k}e^{3imt/4},\quad
    \delta \bar{y} = -\,\bar{\epsilon}^{k}\psi_{k}\,e^{-3imt/4},\nn
    &&\delta \psi^i = \bar{\epsilon}^i\left(2i\dot{z}\right)e^{-3imt/4} + \epsilon^i\left(2i\dot{\bar{y}}-m\bar{y}\right)e^{3imt/4},\nn
    &&\delta \bar{\psi}_i = -\,\epsilon_i\left(2i\dot{\bar{z}}\right)e^{3imt/4}+\bar{\epsilon}_i \left(2i\dot{y}+my\right)e^{-3imt/4},
\eea
where
\bea
    &&\sqrt{2}\,z := \phi\,,\qquad \sqrt{2}\,\bar{z}:=\bar{\phi}\,,\qquad y:=y^{34},\qquad \bar{y}:=y^{12},\nn
    && \psi^1 :=\chi^1,\qquad \psi^2 :=\chi^2,\qquad \bar{\psi}_1 :=\bar{\chi}_1\,,\qquad \bar{\psi}_2 :=\bar{\chi}_2\,.
\eea
These transformations differ from those given in \cite{DHSS}. In the present case, Pauli-G\"ursey ${\rm SU}(2)$ symmetry is broken.
For this case the ${\rm SU}(2|1)$ superfield constraints defining the mirror ${\bf (4, 4, 0)}$ multiplet are written as
\bea
    &&\bar{\cal D}^{i}Z=\bar{\cal D}^{i}Y=0\,,\qquad
    {\cal D}^{i}\bar{Z}={\cal D}^{i}\bar{Y}=0\,,\nn
    &&{\cal D}^{i}Z=-\,\bar{\cal D}^{i}\bar{Y},\qquad {\cal D}^{i}Y=\bar{\cal D}^{i}\bar{Z},\nn
    &&\tilde{F}Z=0\,,\qquad \tilde{F}Y=Y.
\eea
Their solution reads
\bea
    Z &=& z+\theta_{i}\psi^{i}e^{3imt/4}+i\bar{\theta}^{j}\theta_{j} \,\dot{z}
    -\left(\theta\right)^2\left(i\dot{\bar{y}}-\frac{m}{2}\,\bar{y}\right)e^{3imt/2}+\bar{\theta}^{j}\theta_{j}\theta_{i}\left(i\dot{\psi}^{i}-\frac{3m}{4}\,\psi^i\right)e^{3imt/4}\nn
    &&-\,\frac{1}{4}\left(\bar{\theta}\,\right)^2\left(\theta\right)^2\left(\ddot{z} + 2im\dot{z}\right),\nn
    Y &=& y+\theta_{i}\bar{\psi}^{i}e^{3imt/4}+\bar{\theta}^{j}\theta_{j}\left(i\dot{y}+\frac{m}{2}\,y\right)
    +i\left(\theta\right)^2\dot{\bar{z}}\,e^{3imt/2}+\bar{\theta}^{j}\theta_{j} \theta_{i}\left(i\dot{\bar{\psi}}^{i}-\frac{m}{4}\,\bar{\psi}^i\right)e^{3imt/4}
    \nn
    &&-\,\frac{1}{4}\left(\bar{\theta}\,\right)^2\left(\theta\right)^2\left(\ddot{y}+im\dot{y}+\frac{3m^2}{4}\,y\right).\label{YZ}
\eea
\subsubsection{${\rm SU}(2|1)$ superfield action}
The construction of ${\rm SU}(4|1)$ invariant actions in terms of the ${\rm SU}(2|1)$ superfields \eqref{q}, \eqref{YZ} goes as follows.
The general ${\rm SU}(2|1)$ superfield action can be written as
\bea
    S = \int dt\,d^2\theta\,d^2\bar{\theta}\left(1+2m\,\bar{\theta}^k\theta_k\right)f\left(Z,\bar{Z},Y\bar{Y},q^{ia}q_{ia}\right).
\eea
The target space metric $g$ is defined according to \cite{ILS} as
\bea
    &&g=\Delta_2 f= -\,\Delta_1 f\,,\qquad
    f=f\left(z,\bar{z},y\bar{y},x^{ia}x_{ia}\right),\qquad
    g=g\left(z,\bar{z},y\bar{y},x^{ia}x_{ia}\right),\nn
    &&\Delta_1 f + \Delta_2 f = 0 \quad
    \Rightarrow\quad \Delta_1 g + \Delta_2 g = 0\,,\label{Laplace}
\eea
where
\bea
    &&\partial_{ia}=\partial/\partial x^{ia},\qquad \Delta_1 = \varepsilon^{ik}\varepsilon^{ab}\partial_{ia}\partial_{kb}\,,\nn
    &&\partial_z  = \frac{\partial}{\partial z}\,,\quad
    \partial_{\bar{z}} = \frac{\partial}{\partial \bar{z}}\,,\quad
    \partial_y  = \frac{\partial}{\partial y}\,,\quad
    \partial_{\bar{y}} = \frac{\partial}{\partial \bar{y}}\,,\qquad\Delta_2 = 2\left(\partial_z\partial_{\bar{z}}+\partial_y\partial_{\bar{y}}\right).
\eea
Since ${\rm SU}(2|1)$ supersymmetry implies ${\rm SU}(2)\times {\rm U}(1)$ symmetry, the function $f$ and $g$ are functions of the following  coordinate monomials : $z, \bar{z}, y\bar{y}, x^{ia}x_{ia}$\,.

Requiring ${\rm SU}(4)$ invariance of the corresponding component action amounts to the constraints:
\bea
    &&m\left(\bar{y} g + 2\partial_{y}f + x^{ia}\partial_{ia}\partial_{y}f\right) = 0\quad
    \Rightarrow\quad m\left(x_{ia}\partial_y - \bar{y}\partial_{ia}\right)g = 0\,,\nn
    &&m\left(y g + 2\partial_{\bar y}f + x^{ia}\partial_{ia}\partial_{\bar y}f\right) = 0\quad
    \Rightarrow\quad m\left(x_{ia}\partial_{\bar{y}}-y\partial_{ia}\right)g = 0\,.\label{add_cond}
\eea
These constraints admit three different solutions :
\begin{itemize}
\item[1)]Special K\"ahler manifold metric \eqref{SKM}
\bea
     &&f_{1} = \frac{1}{2}\left[\bar{z}\partial_z K\left(z\right)+z\partial_{\bar{z}}\bar{K}\left(\bar{z}\right)\right]-\frac{1}{16}\left(x^{ia}x_{ia}+4y\bar{y}\right)\left[\partial_z \partial_z K\left(z\right)+\partial_{\bar{z}}\partial_{\bar{z}}\bar{K}\left(\bar{z}\right)\right],\nn
    &&g_1 = \frac{1}{2}\left[\partial_z\partial_z K\left(z\right)+\partial_{\bar{z}}\partial_{\bar{z}}\bar{K}\left(\bar{z}\right)\right]\quad \Longrightarrow \quad g_1 = \partial_{\phi}\partial_{\phi} K\left(\phi\right)+\partial_{\bar{\phi}}\partial_{\bar{\phi}}\bar{K}\left(\bar{\phi}\,\right).\label{1}
\eea
\item[2)] ${\rm SO}(6)$-invariant metric \eqref{g_so6}
\bea
    &&f_2 = \frac{1}{4}\left(x^{ia}x_{ia}\right)^{-1}\log{\left(2y\bar{y}+x^{ia}x_{ia}\right)},\nn
    &&g_2 = \left(2y\bar{y}+x^{ia}x_{ia}\right)^{-2}\quad \Longrightarrow \quad g_2 = \left[\frac{1}{2}\,y^{IJ}y_{IJ}\right]^{-2}.\label{2}
\eea
\item[3)] ${\rm SO}(8)$-invariant metric
\bea
    &&f_3 = -\,\frac{1}{8}\left(x^{ia}x_{ia}\right)^{-1}\left(2z\bar{z}+2y\bar{y}+x^{ia}x_{ia}\right)^{-1},\nn
    &&g_3 = \left(2z\bar{z}+2y\bar{y}+x^{ia}x_{ia}\right)^{-3}\quad\Longrightarrow\quad g_3=\left[\phi\bar{\phi}+\frac{1}{2}\,y^{IJ}y_{IJ}\right]^{-3}.\label{3}
\eea
\end{itemize}
The first solution \eqref{1} reproduces the Lagrangian \eqref{L_SKM} with the metric \eqref{SKM}. Other solutions correspond to
new ${\rm SU}(4|1)$ invariant actions.\\

The second solution \eqref{2} gives the Lagrangian
\bea
    {\cal L}_{{\rm SO}(6)} &=& g_2\left[\dot{\phi}\dot{\bar{\phi}}+\frac{1}{2}\,\dot{y}^{IJ}\dot{y}_{IJ}+ \frac{i}{2}\left(\chi^K\dot{\bar{\chi}}_K-\dot{\chi}^K\bar{\chi}_K\right)- \frac{m}{4}\,\chi^K\bar{\chi}_K-\frac{m^2}{8}\,y^{IJ}y_{IJ}\right]\nn
    &&-\,\frac{i}{\sqrt{2}}\,\dot{\bar{\phi}}\,\partial_{IJ}g_2\,\chi^I\chi^J - \frac{i}{\sqrt{2}}\,\dot{\phi}\,\partial^{IJ}g_2\,\bar{\chi}_I\bar{\chi}_J +i\left(\dot{y}_{IK}\,\partial^{JK}g_2-\dot{y}^{JK}\,\partial_{IK}g_2\right)\chi^I\bar{\chi}_J\nn
    &&-\,\frac{1}{2}\,\partial_{IJ}\partial^{KL}g_2\,\chi^I\chi^J\bar{\chi}_K\bar{\chi}_L\,,\label{so6}
\eea
where
\bea
    \partial_{IJ}=\frac{\partial}{\partial y^{IJ}}\,,\qquad \partial_{IJ}y^{KL}=\frac{1}{2}\left(\delta^K_I\delta^L_J-\delta^L_I\delta^K_J\right),\qquad \partial_{IJ}\left(y_{KL}\right)=\frac{1}{2}\,\varepsilon_{IJKL}\,.
\eea
Substitution $i\dot{\phi}=D$ gives ${\rm SU}(4|1)$ invariant Lagrangian for the multiplet ${\bf (6,8,2)}$, which is in fact  superconformal,
with the relevant group ${\rm SU}(4|1,1)$ (see Appendix \ref{682}). \\

The third solution \eqref{3} exhibits an invariance under the maximal $R$-symmetry group ${\rm SO}(8)$ and  produces the component Lagrangian
\bea
    {\cal L}_{{\rm SO}(8)} &=& g_3\left[\dot{\phi}\dot{\bar{\phi}}+\frac{1}{2}\,\dot{y}^{IJ}\dot{y}_{IJ}+ \frac{i}{2}\left(\chi^K\dot{\bar{\chi}}_K-\dot{\chi}^K\bar{\chi}_K\right)+\frac{m}{4}\,\chi^K\bar{\chi}_K-\frac{m^2}{8}\,y^{IJ}y_{IJ}\right]\nn
    &&-\,\frac{i}{\sqrt{2}}\,\dot{\bar{\phi}}\,\partial_{IJ}g_3\,\chi^I\chi^J - \frac{i}{\sqrt{2}}\,\dot{\phi}\,\partial^{IJ}g_3\,\bar{\chi}_I\bar{\chi}_J +i\left(\dot{y}_{IK}\,\partial^{JK}g_3-\dot{y}^{JK}\,\partial_{IK}g_3\right)\chi^I\bar{\chi}_J\nn
    &&+\,\frac{1}{\sqrt{2}}\left(i\dot{y}_{IJ}-\frac{m}{2}\,y_{IJ}\right)\partial_{\phi}g_3\,\chi^I\chi^J+\frac{1}{\sqrt{2}}\left(i\dot{y}^{IJ}+\frac{m}{2}\,y^{IJ}\right)\partial_{\bar{\phi}}g_3\,\bar{\chi}_I\bar{\chi}_J \nn
    && -\, \frac{i}{2}\left(\dot{\phi}\,\partial_{\phi}g_3 - \dot{\bar{\phi}}\,\partial_{\bar{\phi}}g_3\right)\chi^K\bar{\chi}_K-\frac{im}{2}\left(\dot{\phi}\,\bar{\phi} - \dot{\bar{\phi}}\,\phi\right)g_3\nn
    &&+ \,\frac{m}{4}\left(\phi\,\partial_{\phi}g_3 + \bar{\phi}\,\partial_{\bar{\phi}}g_3\right)\chi^K\bar{\chi}_K
    -\frac{1}{\sqrt{2}}\left(\chi^I\chi^J\,\partial_{IJ}\partial_{\phi} g_3 + \bar{\chi}_I\bar{\chi}_J\,\partial^{IJ}\partial_{\bar{\phi}}g_3\right)\chi^K\bar{\chi}_K\nn
        &&-\,\frac{1}{24}\left(\varepsilon_{IJKL}\,\chi^I\chi^J\chi^K\chi^L\,\partial_{\phi}\partial_{\phi} g_3 + \varepsilon^{IJKL}\,\bar{\chi}_I\bar{\chi}_J\bar{\chi}_K\bar{\chi}_L\,\partial_{\bar{\phi}}\partial_{\bar{\phi}}g_3\right)\nn
    &&-\,\frac{1}{2}\,\partial_{IJ}\partial^{KL}g_3\,\chi^I\chi^J\bar{\chi}_K\bar{\chi}_L+\frac{1}{2}\,\partial_{\phi}\partial_{\bar{\phi}}g_3\,\chi^I\bar{\chi}_I\chi^J\bar{\chi}_J\,.\label{L_SO8}
\eea

\subsection{Superconformal symmetry}\label{conf}

Redefining the component fields in \eqref{L_SO8} as
\bea
    &&\phi \rightarrow \phi\,e^{-imt/2},\qquad \chi^I \rightarrow \chi^I\,e^{-imt/4},\nn
    &&\bar{\phi} \rightarrow \bar{\phi}\,e^{imt/2},\qquad \bar{\chi}_I \rightarrow \bar{\chi}_I\,e^{imt/4},
\eea
we eliminate all the deformed terms proportional to $m$ and write the Lagrangian in ${\rm SO}(8)$ invariant formulation:
\bea
    {\cal L}_{\rm conf} &=& g_3\left[\dot{\phi}\dot{\bar{\phi}}+\frac{1}{2}\,\dot{y}^{IJ}\dot{y}_{IJ}+ \frac{i}{2}\left(\chi^K\dot{\bar{\chi}}_K-\dot{\chi}^K\bar{\chi}_K\right)-\frac{m^2}{4}\left(\phi\bar{\phi}+\frac{1}{2}\,y^{IJ}y_{IJ}\right)\right]\nn
    &&-\,\frac{i}{\sqrt{2}}\,\dot{\bar{\phi}}\,\partial_{IJ}g_3\,\chi^I\chi^J - \frac{i}{\sqrt{2}}\,\dot{\phi}\,\partial^{IJ}g_3\,\bar{\chi}_I\bar{\chi}_J +i\left(\dot{y}_{IK}\,\partial^{JK}g_3-\dot{y}^{JK}\,\partial_{IK}g_3\right)\chi^I\bar{\chi}_J\nn
    &&+\,\frac{i}{\sqrt{2}}\left(\dot{y}_{IJ}\,\chi^I\chi^J\,\partial_{\phi}g_3+\dot{y}^{IJ}\bar{\chi}_I\bar{\chi}_J\, \partial_{\bar{\phi}}g_3\right) - \frac{i}{2}\left(\dot{\phi}\,\partial_{\phi}g_3 - \dot{\bar{\phi}}\,\partial_{\bar{\phi}}g_3\right)\chi^K\bar{\chi}_K\nn
    &&-\,\frac{1}{24}\left(\varepsilon_{IJKL}\,\chi^I\chi^J\chi^K\chi^L\,\partial_{\phi}\partial_{\phi} g_3 + \varepsilon^{IJKL}\,\bar{\chi}_I\bar{\chi}_J\bar{\chi}_K\bar{\chi}_L\,\partial_{\bar{\phi}}\partial_{\bar{\phi}}g_3\right)\nn
    &&-\,\frac{1}{\sqrt{2}}\left(\chi^I\chi^J\,\partial_{IJ}\partial_{\phi} g_3 + \bar{\chi}_I\bar{\chi}_J\,\partial^{IJ}\partial_{\bar{\phi}}g_3\right)\chi^K\bar{\chi}_K\nn
    &&-\,\frac{1}{2}\,\partial_{IJ}\partial^{KL}g_3\,\chi^I\chi^J\bar{\chi}_K\bar{\chi}_L+\frac{1}{2}\,
    \partial_{\phi}\partial_{\bar{\phi}}g_3\,\chi^I\bar{\chi}_I\chi^J\bar{\chi}_J\,.\label{L_conf}
\eea
As a result, we obtain ${\rm OSp}(8|2)$ superconformal Lagrangian of the trigonometric type \footnote{Here we follow the terminology suggested in \cite{HT}.}
that contains only $m^2$ terms.
Since the new Lagrangian \eqref{L_conf} is an even function of $m$, it is invariant under two types of ${\rm SU}(4|1)$ transformations, with the deformation parameters $m$ and $-m$\,:
\bea
    &&\delta \phi = -\,\sqrt{2}\,\epsilon_{I}\chi^{I}e^{imt},\qquad \delta \bar{\phi} = \sqrt{2}\,\bar{\epsilon}^{I}\bar{\chi}_{I}\,e^{-imt},\nn
    &&\delta y^{IJ} = -\,2\,\bar{\epsilon}^{[I}\chi^{J]} e^{-imt}+ \varepsilon^{IJKL}\epsilon_{K}\bar{\chi}_{L}\,e^{imt},\nn
    &&\delta \chi^I = \sqrt{2}\,\bar{\epsilon}^I\left(i\dot{\phi}+\frac{m}{2}\,\phi\right)e^{-imt} - 2\,\epsilon_J\left(i\dot{y}^{IJ}-\frac{m}{2}\,y^{IJ}\right)e^{imt},\nn
    &&\delta \bar{\chi}_I = -\,\sqrt{2}\,\epsilon_I\left(i\dot{\bar{\phi}}-\frac{m}{2}\,\bar{\phi}\right)e^{imt} + 2\,\bar{\epsilon}^J\left(i\dot{y}_{IJ}+\frac{m}{2}\,y_{IJ}\right)e^{-imt}, \label{tr_conf11}
\eea
\bea
    &&\delta \phi = -\,\sqrt{2}\,\eta_{I}\chi^{I}e^{-imt},\qquad \delta \bar{\phi} = \sqrt{2}\,\bar{\eta}^{I}\bar{\chi}_{I}\,e^{imt},\nn
    &&\delta y^{IJ} = -\,2\,\bar{\eta}^{[I}\chi^{J]} e^{imt}+ \varepsilon^{IJKL}\eta_{K}\bar{\chi}_{L}\,e^{-imt},\nn
    &&\delta \chi^I = \sqrt{2}\,\bar{\eta}^I\left(i\dot{\phi}-\frac{m}{2}\,\phi\right)e^{imt} - 2\,\eta_J\left(i\dot{y}^{IJ}+\frac{m}{2}\,y^{IJ}\right)e^{-imt},\nn
    &&\delta \bar{\chi}_I = -\,\sqrt{2}\,\eta_I\left(i\dot{\bar{\phi}}+\frac{m}{2}\,\bar{\phi}\right)e^{-imt} + 2\,\bar{\eta}^J\left(i\dot{y}_{IJ}-\frac{m}{2}\,y_{IJ}\right)e^{imt}, \label{tr_conf12}
\eea
In the closure of these transformations,
we obtain superconformal algebra $osp(8|2)$ spanned by 16 supercharges and 31 bosonic generators (see Appendix \ref{osp82}),\footnote{
In the limit $m{=}0$ the Lagrangian \eqref{L_conf} goes into the one
invariant under the ``parabolic''
realization of OSp$(8|2)$, as it was given in \cite{KuTo}.}
where the conformal Hamiltonian ${\cal H}_{\rm conf}$ is defined as
\bea
    {\cal H}_{\rm conf} = {\cal H}-\frac{m}{2}\,F.\label{H_conf}
\eea
The generators $F^{IJ}$ and $\bar{F}_{IJ}$ produce ${\rm SO}(8)/{\rm U}(4)$ transformations realized as
\bea
&&\delta \chi^I = \sqrt{2}\,\bar\Lambda^{IJ}\bar{\chi}_J\,, \qquad \delta \bar{\chi}_I = \sqrt{2}\,\Lambda_{IJ}\chi^J, \nn
&&\delta \phi = -\,\bar\Lambda^{IJ}y_{IJ}\,, \quad  \delta \bar{\phi} = -\,\Lambda^{IJ}y_{IJ}\,,  \quad
\delta y_{IJ} = \Lambda_{IJ}\phi + \bar\Lambda_{IJ}\bar{\phi}.
\eea

\section{The ${\rm SU}(4|1)$ multiplet ${\bf (8,8,0)}$\,: second version}\label{v2}
The second version of the multiplet ${\bf (8,8,0)}$ is described by a complex bosonic superfield $V^I$ satisfying
\bea
    &&{\cal D}^{I}V^{J} = \frac{1}{2}\,\varepsilon^{IJKL}\,\bar{\cal D}_K\bar{V}_L\,,\qquad {\cal D}^{(I}\,V^{J)} = 0\,,\qquad\bar{\cal D}_{(K}\,\bar{V}_{L)}=0\,,\nn
    &&{\cal D}^{I}\bar{V}_J  = \frac{1}{4}\,\delta^I_J {\cal D}^{K}\bar{V}_K \qquad \bar{\cal D}_J\, V^{I}=\frac{1}{4}\,\delta^I_J\bar{\cal D}_K V^{K}\qquad
    \overline{\left(V^I\right)}=\bar{V}_I\,.\label{N8constrv2}
\eea
In the flat superspace limit $m\rightarrow 0$\,, these constraints go over to the ${\rm SU}(4)$ covariant constraints \eqref{constr_v2}
specifying another form of the flat ${\cal N}\,{=}\,8$, $d\,{=}\,1$ multiplet ${\bf (8, 8, 0)}$.

To avoid calculation of the deformed covariant derivatives ${\cal D}^I$ and $\bar{{\cal D}}_J$\,,
 we instead consider harmonization of part of these constraints, {\it viz.}
\bea
\bar{\cal D}_{(K}\,\bar{V}_{L)}=0\,,\qquad {\cal D}^{I}\bar{V}_J  = \frac{1}{4}\,\delta^I_J {\cal D}^{K}\bar{V}_K\,,\label{dconstr}
\eea
 with the rest of constraints being solved at the component level.

\subsection{Harmonic superspace}
 The option for harmonic superspace relevant to the given  case uses the harmonic variables on ${\rm SU}(4)/[{\rm SU}(3)\times {\rm U}(1)]$ \cite{nguyen}.
 The set of these harmonic variables is given by $u^{(+)\alpha}_{I}$, $u^{(+3)I}$, $u^{(-)I}_{\beta}$, $u^{(-3)}_{I}\,,$ where the index $\alpha = 1,2,3$ refers to the ${\rm SU}(3)$ fundamental representation. The harmonics satisfy the following unitarity and unimodularity conditions:
\bea
    &&  u^{(-3)}_I u^{(+3)I} = 1\,,\qquad   u^{(+)\alpha}_I u^{(-)I}_\beta = \delta^{\alpha}_{\beta}\,,\qquad  u^{(+)\alpha}_J u^{(-)I}_\alpha + u^{(-3)}_J u^{(+3)I} = \delta^I_J\,,\nn
    && u^{(+)\alpha}_I u^{(+3)I} = u^{(-3)}_I u^{(-)I}_\alpha  = 0\,,\qquad \varepsilon^{IJKL} u^{(+)\alpha}_I u^{(+)\beta}_J u^{(+)\gamma}_K  u^{(-3)}_L = \varepsilon^{\alpha\beta\gamma}.
\eea
As in the previous case, we define the new coordinates
\bea
    &&\theta^{(+3)} = \theta_{I}\left(u^{(+3)I} + m\,\bar{\theta}^{(+)\alpha}\theta^{(+3)}u^{(-)I}_\alpha\right),\qquad \theta^{(-)}_{\alpha} = \theta_{I}\,u^{(-)I}_\alpha ,\nn
    &&\bar{\theta}^{(+)\alpha} = \bar{\theta}^{J}\left(u^{(+)\alpha}_{J} + m\,\bar{\theta}^{(+)\alpha}\theta^{(+3)}u^{(-3)}_{J}\right),\qquad \bar{\theta}^{(-3)} = \bar{\theta}^{J}u^{(-3)}_{J},\nn
    &&t_{\rm A} = t + i\bar{\theta}^{(-3)} \theta^{(+3)} - i\bar{\theta}^{(+)\alpha}\theta^{(-)}_{\alpha}\left[1-m\,\bar{\theta}^{(+)\beta}\theta^{(-)}_{\beta}+\frac{4m^2}{3}\left(\bar{\theta}^{(+)\beta}\theta^{(-)}_{\beta}\right)^2\right].\label{HSS2}
\eea
They transform as
\bea
    &&\delta \theta^{(-)}_{\alpha} = \epsilon^{(-)}_{\alpha} + 2m\left[\bar{\epsilon}^{(+)\beta}\theta^{(-)}_{\beta}+\bar{\epsilon}^{(-3)}\theta^{(+3)}\left(1+m\,\bar{\theta}^{(+)\beta}\theta^{(-)}_{\beta}\right)\right]\theta^{(-)}_{\alpha},\nn
    &&\delta \bar{\theta}^{(-3)} = \bar{\epsilon}^{(-3)} - 2m\,\epsilon^{(-)}_{\beta}\bar{\theta}^{(+)\beta}\bar{\theta}^{(-3)}, \nn
    &&\delta \theta^{(+3)} = \epsilon^{(+3)} + m\,\epsilon^{(-)}_{\alpha}\bar{\theta}^{(+)\alpha}\theta^{(+3)},\nn
    &&\delta \bar{\theta}^{(+)\alpha} = \bar{\epsilon}^{(+)\alpha} + m\,\bar{\epsilon}^{(-3)}\bar{\theta}^{(+)\alpha}\theta^{(+3)} - 2m\,\epsilon^{(-)}_{\beta}\bar{\theta}^{(+)\beta}\bar{\theta}^{(+)\alpha}, \nn
    &&\delta u^{(+3)I} = \Lambda^{(+4)\alpha}u^{(-)I}_{\alpha}\qquad \delta u^{(-3)}_{I} = 0\,,\nn
    &&\delta u^{(+)\alpha}_I = -\,\Lambda^{(+4)\alpha} u^{(-3)}_{I} \qquad \delta u^{(-)I}_{\beta} = 0\,,\nn
    &&\delta t_A =  2i\left(\epsilon^{(-)}_{\alpha}\bar{\theta}^{(+)\alpha} + \bar{\epsilon}^{(-3)}\theta^{(+3)}\right),\label{trHSS}
\eea
where
\bea
    &&\Lambda^{(+4)\alpha} = m\left(\epsilon^{(+3)}\bar{\theta}^{(+)\alpha} + \bar{\epsilon}^{(+)\alpha}\theta^{(+3)}\right)+m^2\epsilon^{(-)}_{\beta}\bar{\theta}^{(+)\beta}\bar{\theta}^{(+)\alpha}\theta^{(+3)},\nn
    &&\epsilon^{(-)}_{\alpha} = \epsilon_{I}\,u^{(-)I}_\alpha , \quad \epsilon^{(+3)} = \epsilon_{I}\,u^{(+3)I},\quad
    \bar{\epsilon}^{(+)\alpha} = \bar{\epsilon}^{J}u^{(+)\alpha}_{J} , \quad \bar{\epsilon}^{(-3)} = \bar{\epsilon}^{J}u^{(-3)}_J .
\eea
It is straightforward to see that the analytic subspace
\bea
    \zeta_{\rm A} =\left\lbrace t_{\rm A},\theta^{(+3)}, \bar{\theta}^{(+)\alpha},u^{(+)\alpha}_I, u^{(+3)I},u^{(-)I}_{\beta}, u^{(-3)}_{I}\right\rbrace,\label{AHSS}
\eea
is closed under the transformations \eqref{trHSS}. Its integration measure
\bea
    d\zeta_{\rm A}^{(-6)} =  dt_{\rm A}\,du\,d\theta^{(+3)}\,d^3\bar{\theta}^{(+)}\,e^{3imt_{\rm A}/2}
\eea
transforms as
\bea
    \delta\left(d\zeta_{\rm A}^{(-6)}\right) = m\,d\zeta_{\rm A}^{(-6)}\left(\epsilon^{(-)}_{\alpha}\bar{\theta}^{(+)\alpha}-3\,\bar{\epsilon}^{(-3)}\theta^{(+3)}\right).
\eea

The harmonic derivatives are found to be
\bea
    &&{\cal D}^{(+4)\alpha} = \partial^{(+4)\alpha} - 2i\bar{\theta}^{(+)\alpha}\theta^{(+3)}\partial_{\rm A}-\frac{m}{6}\,\bar{\theta}^{(+)\alpha}\theta^{(+3)}{\cal D}^0+m\,\bar{\theta}^{(+)\alpha}\theta^{(+3)}\bar{\theta}^{(+)\beta}\frac{\partial}{\partial \bar{\theta}^{(+)\beta}}\,,\nn
    &&{\cal D}^{\alpha}_{\beta}=\partial^{\alpha}_{\beta}+\bar{\theta}^{(+)\alpha}\frac{\partial}{\partial \bar{\theta}^{(+)\beta}}-\frac{\delta^{\alpha}_{\beta}}{3}\,\bar{\theta}^{(+)\gamma}\frac{\partial}{\partial \bar{\theta}^{(+)\gamma}}\,,\nn
    &&{\cal D}^0 = \partial^0 + \bar{\theta}^{(+)\alpha}\frac{\partial}{\partial \bar{\theta}^{(+)\alpha}}+3\,\theta^{(+3)}\frac{\partial}{\partial \theta^{(+3)}}\,,
\eea
where
\bea
    &&\partial^{(+4)\alpha} = u^{(+3)K}\frac{\partial}{\partial u^{(-)K}_{\alpha}} - u^{(+)\alpha}_K\frac{\partial}{\partial u^{(-3)}_{K}}\,,\nn
    &&\partial^{\alpha}_{\beta} =  u^{(+)\alpha}_K\frac{\partial}{\partial u^{(+)\beta}_K} - u^{(-)K}_{\beta}\frac{\partial}{u^{(-)K}_{\alpha}}-\frac{\delta^{\alpha}_{\beta}}{3}\left(u^{(+)\gamma}_K\frac{\partial}{\partial u^{(+)\gamma}_K} - u^{(-)K}_{\gamma}\frac{\partial}{u^{(-)K}_{\gamma}}\right),\nn
    &&\partial^0 = u^{(+)\alpha}_K\frac{\partial}{\partial u^{(+)\alpha}_K} - u^{(-)K}_{\alpha}\frac{\partial}{u^{(-)K}_{\alpha}}+3\left(u^{(+3)K}\frac{\partial}{\partial u^{(+3)K}}-u^{(-3)}_{K}\frac{\partial}{\partial u^{(-3)}_{K}}\right).
\eea
Note that
\bea
    {\cal D}^{(+4)\alpha}\Lambda = \Lambda^{(+4)\alpha} ,\qquad \Lambda = m\left(\epsilon^{(-)}_{\alpha}\bar{\theta}^{(+)\alpha} - \bar{\epsilon}^{(-3)}\theta^{(+3)}\right).
\eea

The harmonic analytic superfield $\bar{V}^{(+3)}$ defined on \eqref{AHSS} satisfies the harmonic constraints
\bea
    {\cal D}^{(+4)\alpha}\bar{V}^{(+3)} = 0\,,\qquad {\cal D}^{\alpha}_{\beta}\bar{V}^{(+3)} = 0\,,\qquad {\cal D}^{0}\bar{V}^{(+3)} = 3\bar{V}^{(+3)}.\label{constrV+3}
\eea
It can be treated as a harmonization of the superfield $\bar{V}_{I}$ defined by \eqref{dconstr}, where Grassmann analyticity constraints are provided by
\bea
u^{(+3)K}u^{(+3)L}\bar{\cal D}_{(K}\,\bar{V}_{L)}=\bar{\cal D}^{(+3)}\,\bar{V}^{(+3)}=0\,,\quad u^{(+)\alpha}_{I}u^{(+3)J}{\cal D}^{I}\bar{V}_J  =  {\cal D}^{(+)\alpha}\bar{V}^{(+3)}=0\,.
\eea
The full set of the constraints \eqref{N8constrv2} operates with the set of superfields $V^{(+)\alpha}$, $V^{(-3)}$, $\bar{V}^{(+3)}$ and $\bar{V}^{(-)}_{\alpha}$ living on the full harmonic superspace \eqref{HSS2}\,. Here we consider just the superfield $\bar{V}^{(+3)}$ treated as an unconstrained deformed harmonic superfield satisfying
the analyticity conditions \eqref{constrV+3}\,. The  rest of constraints on $\bar{V}^{(+3)}$ will be imposed below ``by hand''  at the component level, like in the previous cases.

The general expansion of $\bar{V}^{(+3)}$ reads
\bea
    \bar{V}^{(+3)} &=& \bar{z}_I\,u^{(+3)I} + \sqrt{2}\,\theta^{(+3)}\chi\,e^{3imt/4} + 2\,\bar{\theta}^{(+)\alpha}\chi_{JI}\,u^{(+3)J} u^{(-)I}_\alpha\,e^{-3imt/4}
    \nn
    &&+\,2\,\bar{\theta}^{(+)\alpha}\theta^{(+3)}\left(i\dot{\bar{z}}_K+\frac{m}{4}\,\bar{z}_K\right)u^{(-)K}_{\alpha}\nn
    &&+\,\varepsilon_{IJKL}\,\bar{\theta}^{(+)\alpha}\bar{\theta}^{(+)\beta}C^K u^{(+3)L}u^{(-)I}_{\alpha}u^{(-)J}_{\beta}e^{-3imt/2}\nn
    &&+\,\varepsilon_{\alpha\beta\gamma}\,\bar{\theta}^{(+)\alpha}\bar{\theta}^{(+)\beta}\bar{\theta}^{(+)\gamma}\pi\,e^{-9imt/4}\nn
    &&-\,2\,\bar{\theta}^{(+)\alpha}\bar{\theta}^{(+)\beta}\theta^{(+3)}\left(i\dot{\chi}_{IJ}+\frac{m}{2}\,{\chi}_{IJ}\right)u^{(-)I}_{\alpha}u^{(-)J}_{\beta}e^{-3imt/4}\nn
    &&-\,\frac{2}{3}\,\varepsilon_{IJKL}\,\bar{\theta}^{(+)\alpha}\bar{\theta}^{(+)\beta}\bar{\theta}^{(+)\gamma}\theta^{(+3)}\left(i\dot{C}^{L}+\frac{3m}{4}\,C^{L}\right)u^{(-)I}_{\alpha}u^{(-)J}_{\beta}u^{(-)K}_{\gamma}\,e^{-3imt/2},\nn
\eea
where
\bea
    \chi_{IJ}\equiv \chi_{[IJ]}\,.
\eea
Taking into account the transformation rule
\bea
    \delta {\cal D}^{(+4)\alpha} &=& -\left(\frac{1}{3}\,\Lambda^{(+4)\alpha} {\cal D}^0 + \Lambda^{(+4)\beta}{\cal D}^{\alpha}_{\beta}\right)
    -\frac{m}{6}\left(\bar{\epsilon}^{(+)\alpha}\theta^{(+3)}-\epsilon^{(+3)}\bar{\theta}^{(+)\alpha}\right){\cal D}^0\nn
    &&+\,\frac{m^2}{6}\,\epsilon^{(-)}_{\beta}\bar{\theta}^{(+)\beta}\bar{\theta}^{(+)\alpha}\theta^{(+3)}{\cal D}^0,
\eea
the superfield $\bar{V}^{(+3)}$ transforms as
\bea
    \delta \bar{V}^{(+3)} = \Lambda \bar{V}^{(+3)}-\frac{m}{2}\left(\epsilon^{(-)}_{\alpha}\bar{\theta}^{(+)\alpha}+\bar{\epsilon}^{(-3)}\theta^{(+3)}\right)\bar{V}^{(+3)}.
\eea
This superfield transformation law amounts to the following component transformations
\bea
    &&\delta \bar{z}_{J} = -\,2\,\bar{\epsilon}^K\chi_{JK}\,e^{-3imt/4}-\sqrt{2}\,\epsilon_{J}\chi\,e^{3imt/4},\nn
    &&\delta \chi = \sqrt{2}\,\bar{\epsilon}^K\left(i\dot{\bar z}_K +\frac{3m}{4}\,\bar{z}_K\right)\,e^{-3imt/4},\nn
    &&\delta \chi_{IJ} = \varepsilon_{IJKL}\,\bar{\epsilon}^{K}C^L\,e^{-3imt/4} - 2\,\epsilon_{[I}\left( i\dot{\bar z}_{J]} - \frac{m}{4}\,\bar{z}_{J]}\right)e^{3imt/4},\nn
    &&\delta C^{I} = \varepsilon^{IJKL}\epsilon_J\left(i\dot{\chi}_{KL} - \frac{m}{2}\,\chi_{KL}\right)e^{3imt/4}-3\,\bar{\epsilon}^I \pi\,e^{-3imt/4},\nn
    &&\delta \pi = \frac{2}{3}\,\epsilon_K\left(i\dot{C}^{K}-\frac{3m}{4}\,C^{K}\right)e^{3imt/4}.\label{tr_V+3}
\eea
{}From the transformation properties of $\bar{V}^{(+3)}$ one can draw the conclusion that the construction of a ``pre-action'' similar to \eqref{componentChir}  cannot
be performed within the analytic harmonic superspace. We conjecture that such a construction could become  possible after taking account of the additional set of c
constraints defining the multiplet ${\bf (8,8,0)}$. Then the action can probably be constructed in the full harmonic superspace approach (see \cite{IvSmi}).

At the component level, the rest of the constraints \eqref{N8constrv2} impose the relations
\bea
    && \overline{\left(\bar{z}_I\right)}=z^I,\qquad \overline{\left(\chi\right)}=\bar{\chi}\,,\qquad \overline{\left(\chi_{IJ}\right)}=\frac{1}{2}\,\varepsilon^{IJKL}\chi_{KL}=\chi^{IJ},\nn
    && C^{I} = i\dot{z}^I + \frac{m}{4}\,z^I,\qquad
    \pi = -\,\frac{\sqrt{2}}{3}\left(i\dot{\bar\chi} + m\,\bar{\chi}\right).\label{add_v2}
\eea
The final form of the deformed transformations is
\bea
    &&\delta z^{I} = 2\,\epsilon_K\chi^{IK}\,e^{3imt/4}+\sqrt{2}\,\bar{\epsilon}^{I}\bar{\chi}\,e^{-3imt/4}\,,\nn
    &&\delta \bar{z}_{J} = -\,2\,\bar{\epsilon}^K\chi_{JK}\,e^{-3imt/4}-\sqrt{2}\,\epsilon_{J}\chi\,e^{3imt/4},\nn
    &&\delta \chi = \sqrt{2}\,\bar{\epsilon}^K\left(i\dot{\bar z}_K + \frac{3m}{4}\,\bar{z}_K\right)e^{-3imt/4},\nn
    &&\delta \bar{\chi} = -\,\sqrt{2}\,\epsilon_K\left(i\dot{z}^K - \frac{3m}{4}\,z^K\right)e^{3imt/4},\nn
    &&\delta \chi^{IJ} = 2\,\bar{\epsilon}^{[I}\left(i\dot{z}^{J]} + \frac{m}{4}\,z^{J]}\right)e^{-3imt/4} - \varepsilon^{IJKL}\epsilon_K\left(i\dot{\bar{z}}_L - \frac{m}{4}\,\bar{z}_L\right)e^{3imt/4},
\eea
where
\bea
    \overline{\left(z^{I}\right)}=\bar{z}_{I}\,,\qquad
    \overline{\left(\chi\right)}=\bar{\chi}\,,\qquad
    \overline{\left(\chi^{IJ}\right)}=\chi_{IJ}=\frac{1}{2}\,\varepsilon_{IJKL}\,\chi^{KL}.
\eea

\subsection{${\rm SU}(2|1)$ superfield formulation}
Once again, we split the given multiplet into ${\rm SU}(2|1)$ multiplets as ${\bf (4,4,0)}\oplus{\bf (4,4,0)}$.

The first multiplet is associated with the fields
\bea
    x^{i1}:=z^i,\quad x^{2}_i:=\bar{z}_i\,,\qquad
    \xi^{1}:=2\chi^{12},\quad \bar{\xi}_{1}:=2\chi_{12}\,,\quad
    \xi^{2}:=\sqrt{2}\,\chi\,,\quad \bar{\xi}_{2}:=\sqrt{2}\,\bar{\chi}\,,
\eea
such that
\bea
    &&\delta x^{iA} =-\,\epsilon^i\xi^A\,e^{3imt/4}  -\bar{\epsilon}^i \bar{\xi}^A\,e^{-3imt/4},\nn
    &&\delta \xi^1 = 2\,\bar{\epsilon}^k \left(i\dot{x}^{1}_{k}+\frac{m}{4}\,x^{1}_k\right)e^{-3imt/4},\quad
    \delta \bar{\xi}_1 = 2\,\epsilon_k \left(i\dot{x}^{k}_{1}-\frac{m}{4}\,x^{k}_{1}\right)e^{3imt/4},\nn
    &&\delta \xi^2 = 2\,\bar{\epsilon}^k \left(i\dot{x}^{2}_{k}-\frac{3m}{4}\,x^{2}_k\right)e^{-3imt/4},\quad
    \delta \bar{\xi}_2 = 2\,\epsilon_k \left(i\dot{x}^{k}_{2}+\frac{3m}{4}\,x^{k}_{2}\right)e^{3imt/4}.
\eea
This first multiplet ${\bf (4, 4, 0)}$ is accommodated  by a superfield $q^{iA}$ obeying the ${\rm SU}(2|1)$ covariant constraints
\bea
    {\cal D}^{(k}q^{i)A} =0\,,\quad \bar{{\cal D}}^{(k}q^{i)A} =0\,, \qquad \tilde{F}q^{iA} = -\frac{1}{2}\left(\sigma_3\right)^{A}_{B}q^{iB}\,, \qquad \overline{\left(q^{iA}\right)}=q_{iA}\,.
\eea
As distinct from \eqref{440constr1}, Pauli-G\"ursey ${\rm SU}(2)$ symmetry is broken.
Taking into account \eqref{derivatives}, we solve these constraints as
\bea
    q^{iA}&=&\left[1+ \frac{m}{2}\,\bar{\theta}^k\theta_k - \frac{5m^2}{16}\left(\bar{\theta}\,\right)^2\left(\theta\right)^2\right]x^{iA}
    -i\varepsilon_{kl}\left(\bar{\theta}^i\theta^l + \bar{\theta}^l\theta^i\right)\left(\dot{x}^{kA}+\frac{im}{4}\left(\sigma_3\right)^{A}_{B}x^{kB}\right)\nn
    &&-\,i\bar{\theta}^k\theta_k\left(\theta^i \dot{\xi}^A\,e^{3imt/4}-\bar{\theta}^i\dot{\bar{\xi}}^A\,e^{-3imt/4}\right)\nn
    &&+\left(1+\frac{m}{4}\,\bar{\theta}^k\theta_k\right)\left(\theta^i\xi^A\,e^{3imt/4} + \bar{\theta}^i\bar{\xi}^A\,e^{-3imt/4}\right)\nn
    &&+\,\frac{m}{4}\,\bar{\theta}^k\theta_k\left(\theta^i\xi^B\,e^{3imt/4} - \bar{\theta}^i\bar{\xi}^B\,e^{-3imt/4}\right)\left(\sigma_3\right)^{A}_{B}\nn
    &&+\,\frac{1}{4}\left(\bar{\theta}\,\right)^2\left(\theta\right)^2\left(\ddot{x}^{iA}+\frac{im}{2}\left(\sigma_3\right)^{A}_{B}\dot{x}^{iB} - \frac{m^2}{16}\,x^{iA}\right).
\eea

The second (mirror) multiplet ${\bf (4,4,0)}$ is formed by the fields
\bea
    y^1 := z^4,\quad y^2:=z^3,\quad \bar{y}_1:=\bar{z}_4\,,\quad \bar{y}_2:=\bar{z}_3\,,\qquad
    \psi^{i1}:=2\chi^{i4},\quad \psi^{i2}:=2\chi^{i3},
\eea
with the ${\rm SU}(2|1)$ transformations
\bea
    &&\delta y^{a} =-\,\epsilon_i\psi^{ia}\,e^{3imt/4},\qquad    \delta \bar{y}^a =-\,\bar{\epsilon}_i \psi^{ia}\,e^{-3imt/4},\nn
    &&\delta \psi^{ia} =2\,\bar{\epsilon}^i\left(i\dot{y}^{a} +\frac{m}{4}\,y^{a}\right)e^{-3imt/4}-2\,\epsilon^i \left(i\dot{\bar{y}}^{a} - \frac{m}{4}\,\bar{y}^{a}\right)e^{3imt/4}.
\eea
The superfield ${\rm SU}(2|1)$ constraints defining the mirror ${\bf (4, 4, 0)}$ multiplet
are written as
\bea
    &&\bar{\cal D}^{i} Y^{a} ={\cal D}^{i} \bar{Y}^{a} = 0\,,
    \qquad {\cal D}^{i} Y^{a}=\bar{\cal D}^{i}\bar{Y}^{a}, \nn
    &&\tilde{F}Y^a = \frac{1}{2}\,Y^a,\qquad \tilde{F}\bar{Y}^a = -\,\frac{1}{2}\,\bar{Y}^a,\qquad \overline{\left(Y^a\right)}=\bar{Y}_a\,.
\eea
They are  solved by
\bea
    Y^{a} &=& \left[1+\frac{m}{4}\,\bar{\theta}^{k}\theta_{k} - \frac{7m^2}{64}\left(\bar{\theta}\,\right)^2\left(\theta\right)^2\right] y^{a}
    + i\dot{y}^{a}\left[\bar{\theta}^{k}\theta_{k} - \frac{3m}{8}\left(\bar{\theta}\,\right)^2\left(\theta\right)^2\right]-\frac{1}{4}\left(\bar{\theta}\,\right)^2\left(\theta\right)^2\ddot{y}^a\nn
    &&+\,\theta_{k}\theta^{k}\left(i\dot{\bar{y}}^{a} - \frac{m}{4}\,\bar{y}^{a}\right)\,e^{3imt/2} + \left[\left(1-\frac{m}{2}\,\bar{\theta}^{k}\theta_{k}\right)\theta_{i} \psi^{ia}+ i\bar{\theta}^{k}\theta_{k} \theta_{i}\dot{\psi}^{ia}\right]e^{3imt/4} ,\nn
    \bar{Y}^{a} &=& \left[1+\frac{m}{4}\,\bar{\theta}^{k}\theta_{k} - \frac{7m^2}{64}\left(\bar{\theta}\,\right)^2\left(\theta\right)^2\right]\bar{y}^{a}
     - i\dot{\bar{y}}^{a}\left[\bar{\theta}^{k}\theta_{k} - \frac{3m}{8}\left(\bar{\theta}\,\right)^2\left(\theta\right)^2\right]-\frac{1}{4}\left(\bar{\theta}\,\right)^2\left(\theta\right)^2 \ddot{\bar{y}}^a
    \nn
    &&+ \,\bar{\theta}^{k} \bar{\theta}_{k} \left( i\dot{y}^{a} + \frac{m}{4}\,y^{a}\right)\,e^{-3imt/2}+\left[\left(1-\frac{m}{2}\,\bar{\theta}^{k}\theta_{k}\right) \bar{\theta}_{i}\psi^{ia}-i\bar{\theta}^{k}\theta_{k}\bar{\theta}_{i}\dot{\psi}^{ia}\right]e^{-3imt/4}.\nn
\eea

\subsection{Invariant Lagrangian}
The general ${\rm SU}(2|1)$ invariant action is written as
\bea
    S = \int dt\,{\cal L} = \frac{1}{2}\int dt\,d^2\theta\,d^2\bar{\theta}\left(1+2m\,\bar{\theta}^k\theta_k\right)f\left(Y^a\bar{Y}_a,q^{iA}q_{iA}\right).
\eea
Requiring it to be ${\rm SU}(4)$ invariant produces the following conditions:
\bea
    &&\Delta_y  = -\, 2\,\varepsilon^{ab}\partial_{a}\bar{\partial}_{b}\,,\qquad \partial_a
    = \partial/\partial y^a,\quad
    \bar{\partial}_b = \partial/\partial \bar{y}^b,\nn
    &&\Delta_x = \varepsilon^{ij}\varepsilon^{AB}\partial_{iA}\partial_{jB}\,,\qquad \partial_{iA}=\partial/\partial x^{iA},\nn
    &&G:=\Delta_y f = -\,\Delta_x f\quad \Rightarrow \quad \left(\Delta_y + \Delta_x\right)G=0\,,
\eea
\bea
    &&m\left(2\partial_a f + \bar{y}_{a}G + x^{iA}\partial_{iA}\partial_a f\right) = 0\quad \Rightarrow \quad
    m\left(\bar{y}_{a}\partial_{iA} - x_{iA}\partial_a\right)G = 0\,,\nn
    &&m\left(2\bar{\partial}_a f - y_{a}G + x^{iA}\partial_{iA}\bar{\partial}_a f\right) = 0\quad \Rightarrow \quad
    m\left(y_{a}\partial_{iA} + x_{iA}\bar{\partial}_a\right)G = 0\,.
\eea
The unique solution of these equations is given by
\bea
    &&f=\frac{1}{4}\left(y^a\bar{y}_a\right)^{-1}\left(y^a\bar{y}_a+\frac{1}{2}\,x^{iA}x_{iA}\right)^{-1}+c_1\left(y^a\bar{y}_a\right)^{-1}\left(x^{iA}x_{iA}\right)^{-1}+ c_2\left(x^{iA}x_{iA}\right)^{-1}\quad \Rightarrow\nn &&\quad \Rightarrow\quad G = \left(y^a\bar{y}_a+\frac{1}{2}\,x^{iA}x_{iA}\right)^{-3}.
\eea
Here, the terms with the constants $c_1$ and $c_2$  do not affect the metric $G$, since it is a harmonic function.
One can check that these terms drop out of the component Lagrangian, which is finally written as
\bea
    {\cal L} &=& \left[\dot{z}^{I}\dot{\bar{z}}_{I}+\frac{i}{2}\,\chi^{IJ}\dot{\chi}_{IJ}+\frac{i}{2}\left(\chi\dot{\bar{\chi}}-\dot{\chi}\bar{\chi}\right)- \frac{im}{4}\left(\dot{z}^{I}\bar{z}_{I}-z^{I}\dot{\bar{z}}_{I}\right)+\frac{m}{4}\,\chi\bar{\chi}-\frac{3m^2}{16}\,{z}^{I}{\bar{z}}_{I}\right]G\nn
    &&+\,i\left(\dot{z}^{I}\partial_J G - \dot{\bar z}_{J}\bar{\partial}^{I}G\right)\chi_{IK}\chi^{JK}-\frac{m}{4}\left(z^{I}\partial_J G + \bar{z}_{J}\bar{\partial}^{I}G\right)\chi_{IK}\chi^{JK}\nn
    &&+\,\frac{i}{2}\left(\dot{z}^{I}\partial_I G - \dot{\bar z}_{I}\bar{\partial}^{I}G\right)\chi\bar{\chi}+\partial_J\bar{\partial}^I G\,\chi_{IK}\chi^{JK}\chi\bar\chi + \frac{1}{3}\,\partial_J\bar{\partial}^I G\,\chi_{IK}\chi^{LK}\chi_{LM}\chi^{JM}\nn
    &&-\,\frac{\sqrt{2}}{3}\,\partial_I\partial_J G\,\bar{\chi}\,\chi^{IK}\chi^{JL}\chi_{KL}
    -\frac{\sqrt{2}}{3}\,\bar{\partial}^I\bar{\partial}^J G\,\chi\,\chi_{IK}\chi_{JL}\chi^{KL},\label{L880_2}
\eea
where
\bea
    G = \left(z^I\bar{z}_I\right)^{-3}.
\eea
\subsection{Superconformal symmetry}\label{conf2}
By analogy with the Section \ref{conf}, one can redefine the component fields as
\bea
    z^I \rightarrow z^I e^{-imt/4},\qquad \bar{z}_I \rightarrow \bar{z}_I\,e^{imt/4},\qquad
    \chi \rightarrow \chi\, e^{imt/2},\qquad \bar{\chi} \rightarrow \bar{\chi}\,e^{-imt/2},
\eea
after which the Lagrangian \eqref{L880_2} becomes an even function of $m$. As a result, we obtain ${\rm OSp}(8|2)$ superconformal Lagrangian that is equivalent to \eqref{L_conf}:
\bea
    {\cal L}_{\rm conf} &=& \left[\dot{z}^{I}\dot{\bar{z}}_{I}+\frac{i}{2}\,\chi^{IJ}\dot{\chi}_{IJ}+\frac{i}{2}\left(\chi\dot{\bar{\chi}}-\dot{\chi}\bar{\chi}\right)-\frac{m^2}{4}\,{z}^{I}{\bar{z}}_{I}\right]G
    +i\left(\dot{z}^{I}\partial_J G - \dot{\bar z}_{J}\bar{\partial}^{I}G\right)\chi_{IK}\chi^{JK}\nn
    &&+\,\frac{i}{2}\left(\dot{z}^{I}\partial_I G - \dot{\bar z}_{I}\bar{\partial}^{I}G\right)\chi\bar{\chi}+\partial_J\bar{\partial}^I G\,\chi_{IK}\chi^{JK}\chi\bar\chi + \frac{1}{3}\,\partial_J\bar{\partial}^I G\,\chi_{IK}\chi^{LK}\chi_{LM}\chi^{JM}\nn
    &&-\,\frac{\sqrt{2}}{3}\,\partial_I\partial_J G\,\bar{\chi}\,\chi^{IK}\chi^{JL}\chi_{KL}
    -\frac{\sqrt{2}}{3}\,\bar{\partial}^I\bar{\partial}^J G\,\chi\,\chi_{IK}\chi_{JL}\chi^{KL}.\label{L_conf2}
\eea
In the same way, this Lagrangian is invariant under two types of $\epsilon_I$ and $\eta_I$ transformations which close on the superalgebra $osp(8|2)$ \eqref{conf_algebra_osp82_ff} - \eqref{conf_algebra_osp82_bf}\,:
\bea
    &&\delta z^{I} = 2\,\epsilon_K\chi^{IK}\,e^{imt}+\sqrt{2}\,\bar{\epsilon}^{I}\bar{\chi}\,e^{-imt},\qquad
    \delta \bar{z}_{J} = -\,2\,\bar{\epsilon}^K\chi_{JK}\,e^{-imt}-\sqrt{2}\,\epsilon_{J}\chi\, e^{imt},\nn
    &&\delta \chi = \sqrt{2}\,\bar{\epsilon}^K\left(i\dot{\bar z}_K + \frac{m}{2}\,\bar{z}_K\right)e^{-imt},\qquad
    \delta \bar{\chi} = -\,\sqrt{2}\,\epsilon_K\left(i\dot{z}^K - \frac{m}{2}\,z^K\right)e^{imt},\nn
    &&\delta \chi^{IJ} = 2\,\bar{\epsilon}^{[I}\left(i\dot{z}^{J]} + \frac{m}{2}\,z^{J]}\right)e^{-imt} - \varepsilon^{IJKL}\epsilon_K\left(i\dot{\bar{z}}_L-\frac{m}{2}\,\bar{z}_L\right)e^{imt},\label{tr_conf1}
\eea
\bea
    &&\delta z^{I} = 2\,\eta_K\chi^{IK}\,e^{-imt}+\sqrt{2}\,\bar{\eta}^{I}\bar{\chi}\,e^{imt},\qquad
    \delta \bar{z}_{J} = -\,2\,\bar{\eta}^K\chi_{JK}\,e^{imt}-\sqrt{2}\,\eta_{J}\chi\, e^{-imt},\nn
    &&\delta \chi = \sqrt{2}\,\bar{\eta}^K\left(i\dot{\bar z}_K - \frac{m}{2}\,\bar{z}_K\right)e^{imt},\qquad
    \delta \bar{\chi} = -\,\sqrt{2}\,\eta_K\left(i\dot{z}^K + \frac{m}{2}\,z^K\right)e^{-imt},\nn
    &&\delta \chi^{IJ} = 2\,\bar{\eta}^{[I}\left(i\dot{z}^{J]} - \frac{m}{2}\,z^{J]}\right)e^{imt} - \varepsilon^{IJKL}\eta_K\left(i\dot{\bar{z}}_L+\frac{m}{2}\,\bar{z}_L\right)e^{-imt}.\label{tr_conf2}
\eea

We see that the Lagrangians \eqref{L_conf} and \eqref{L_conf2} have conformally flat metrics $g_3$ and $G$
which both depend on the quadratic SO(8) invariants of the same power $-3$. 
The fields $z^{I}$ and $\bar{z}_{J}$ can be reexpressed, by a linear transformation,
through the bosonic fields $y^{I^\prime J^\prime}$, $\phi$ and $\bar{\phi}$ of the first multiplet ${\bf (8,8,0)}$,
where $I^\prime$ and $J^\prime$ label the fundamental representation of a different SU$(4)^{\prime}$ subgroup of the SO(8) symmetry, such that
it intersects with the first SU(4) in a common SU(3) subgroup. After  an analogous linear transformation of the fermionic fields, the
Lagrangian \eqref{L_conf2} will coincide with \eqref{L_conf}. So both superconformal Lagrangians are indeed equivalent.
This feature of equivalence of ${\bf (8,8,0)}$ multiplets in the presence of exact SO(8) symmetry was already noted in the end of Section~3.

\section{Summary and outlook}
We have shown the existence of two non-equivalent ``root'' multiplets ${\bf (8, 8, 0)}$
of deformed ${\cal N}\,{=}\,8$ supersymmetry associated
with the supergroup ${\rm SU}(4|1)$. We described them in multiple ways
with worldline superfields and in components and derived invariant actions for them.
Some of these actions are superconformally ${\rm OSp}(8|2)$ invariant.
For a non-trivially interacting example we gave the explicit form of the (classical)
${\rm SU}(4|1)$ supercharges.
We also obtained the ${\rm SU}(4|1)$ invariant actions for the off-shell multiplets
${\bf (6, 8, 2)}$ and ${\bf (7, 8, 1)}$ (in Appendices B and C) from the ${\bf (8, 8, 0)}$ actions,
reconfirming the root interpretation of the ${\bf (8, 8, 0)}$ multiplets for ${\cal N}\,{=}\,8$ mechanics.
The ${\bf (6, 8, 2)}$ action was shown to exhibit superconformal ${\rm SU}(4|1,1)$ invariance.

As for further applications of these results, the most appropriate arena might be provided
by supersymmetric matrix models (see, e.g., \cite{BMN}, \cite{DSV1}, \cite{DSV2}).
These possess ${\rm SU}(4|2)$ invariance, hence multi-particle mechanics based on
${\rm SU}(2|2) \subset {\rm SU}(4|2)$ or ${\rm SU}(4|1) \subset {\rm SU}(4|2)$
may appear as some truncation of such matrix models.
The matrix models studied so far lead to free worldline multiplets and actions.
Our approach allows one to generate non-trivial interactions, which hopefully
may be interpreted as effective actions with quantum corrections taken into account.
An important ingredient of matrix models is a gauging of appropriate isometries
by non-propagating gauge multiplets.
To promote this to the ${\rm SU}(4|1)$ superfield language, one needs to define
suitable gauge superfields generalizing those used in \cite{DI}, \cite{FIL08} or~\cite{FI16}.

Another problem for the future is finding an action including both types of deformed
${\bf (8, 8, 0)}$ multiplets and inquiring the ensuing target-space geometry.

\section*{Acknowledgements}
\noindent We thank Sergey Fedoruk, Armen Nersessian and Francesco Toppan for
interest in this work and for valuable comments. This research was
supported by the Heisenberg-Landau program and the joint DFG project
LE 838/12. The work of  E.I. and S.S. was supported by the Russian
Foundation for Basic Research, project No.~18-02-01046. They thank
the directorate of the Institute of Theoretical Physics at Leibniz
University of Hannover for kind hospitality extended to them several times
during this project.

\appendix
\section{Some calculations}\label{calc}
Here we collect the necessary identities for calculation of the function $L^{(+4)}$ in the invariant action \eqref{L+4act}. We represent it as an infinite series:
\bea
    L^{(+4)}=\sum_{n=0}^{\infty} a_n\left(c^{(-2)}\right)^{n}\left(\hat{Y}^{(+2)}\right)^{n+2}+m^2\left(\theta^{(+)}\right)^4\sum_{n=1}^{\infty} b_n\left(c^{(-2)}\hat{Y}^{(+2)}\right)^{n}.\label{series}
\eea
All the identities below are given up to terms with a total harmonic derivative ${\cal D}^{(+2)i}_{a}$. In addition, one must take into account the definitions \eqref{hatY}.
Each term in the variation of the series \eqref{series} also contains transformations compensating the measure transformations \eqref{measure2}, {\it i.e.}
\bea
    \delta\left(\hat{Y}^{(+2)}\right)^{2}&=&
    4\Lambda^{(0)}\left(\hat{Y}^{(+2)}\right)^{2}+2\Lambda^{(0)}c^{(+2)}\hat{Y}^{(+2)}
    +2\,\varepsilon^{ab}\varepsilon_{ij}\,\Lambda^{(+2)i}_{a}{\cal D}^{(+2)j}_{b}c^{(-2)}\hat{Y}^{(+2)}\nn
    &&-\,\frac{1}{2}\,\varepsilon^{ab}\varepsilon_{ij}\,c^{(-2)}\hat{Y}^{(+2)}\left({\cal D}^{(+2)j}_{b}\Lambda^{(+2)i}_{a}\right),
\eea
\bea
    \delta\left(c^{(-2)}\right)^n\left(\hat{Y}^{(+2)}\right)^{n+2}
    &=&\left(n+4\right)\Lambda^{(0)}\left(c^{(-2)}\right)^n\left(\hat{Y}^{(+2)}\right)^{n+2}\nn &&+\left(n+2\right)\Lambda^{(0)}c^{(+2)}\left(c^{(-2)}\right)^n\left(\hat{Y}^{(+2)}\right)^{n+1}\nn
    &&+\,\frac{\left(n+2\right)}{\left(n+1\right)}\,\varepsilon^{ab}\varepsilon_{ij}\,\Lambda^{(+2)i}_{a}{\cal D}^{(+2)j}_{b}\left(c^{(-2)}\right)^{n+1}\left(\hat{Y}^{(+2)}\right)^{n+1}\nn
    &&-\,\frac{\left(n+2\right)}{4}\,\varepsilon^{ab}\varepsilon_{ij}\,\left({\cal D}^{(+2)j}_{b}\Lambda^{(+2)i}_{a}\right)\left(c^{(-2)}\right)^{n+1}\left(\hat{Y}^{(+2)}\right)^{n+1}\nn
    &=&\left(n+4\right)\Lambda^{(0)}\left(c^{(-2)}\right)^n\left(\hat{Y}^{(+2)}\right)^{n+2}\nn
    &&+\,n\,\Lambda^{(0)}\left(c^{(-2)}\right)^{n-1}\left(\hat{Y}^{(+2)}\right)^{n+1}\nn
    &&-\,\frac{\left(n+3\right)\left(n+4\right)}{4\left(n+1\right)}\,\varepsilon^{ab}\varepsilon_{ij}\left({\cal D}^{(+2)j}_{b}\Lambda^{(+2)i}_{a}\right)\left(c^{(-2)}\hat{Y}^{(+2)}\right)^{n+1}
\eea
where
\bea
    \Lambda^{(0)}\left(c^{(-2)}\right)^nc^{(+2)}&=&-\,\frac{1}{4}\,\Lambda^{(0)}\left(c^{(-2)}\right)^n\left(\varepsilon^{ab}\varepsilon_{ij}\,{\cal D}^{(+2)i}_{a}{\cal D}^{(+2)j}_{b}c^{(-2)}\right)\nn
    &=&\frac{1}{4}\,\Lambda^{(+2)i}_{a}\left(c^{(-2)}\right)^n\left(\varepsilon^{ab}\varepsilon_{ij}\,{\cal D}^{(+2)j}_{b}c^{(-2)}\right) \nn
    &&+\,\frac{n}{4}\,\Lambda^{(0)}\left(c^{(-2)}\right)^{n-1}\left(\varepsilon^{ab}\varepsilon_{ij}\,{\cal D}^{(+2)j}_{b}c^{(-2)}\right)\left({\cal D}^{(+2)i}_{a}c^{(-2)}\right)\nn
    &=&\frac{1}{4\left(n+1\right)}\,\Lambda^{(+2)i}_{a}\,\varepsilon^{ab}\varepsilon_{ij}\,{\cal D}^{(+2)j}_{b}\left(c^{(-2)}\right)^{n+1} \nn
    &&+\,\frac{n}{4}\,\Lambda^{(0)}\left(c^{(-2)}\right)^{n-1}\left(\varepsilon^{ab}\varepsilon_{ij}\,{\cal D}^{(+2)j}_{b}c^{(-2)}\right)\left({\cal D}^{(+2)i}_{a}c^{(-2)}\right).
\eea
We also use the identity ($c^{IJ}c_{IJ} = 4$)
\bea
    \Lambda^{(0)}\left[c^{(-2)}c^{(+2)}+\frac{1}{2}\,\varepsilon^{ab}\varepsilon_{ij}\left({\cal D}^{(+2)j}_{b}c^{(-2)}\right)\left({\cal D}^{(+2)i}_{a}c^{(-2)}\right)\right]=\frac{1}{4}\,c^{IJ}c_{IJ}\,\Lambda^{(0)}=\Lambda^{(0)}\,.
\eea
Thus, the series is invariant when
\bea
    &&a_0 = 1\,,\qquad a_1 = -\,4a_0\,,\qquad a_2=\frac{a_0\,5!}{2!\,3!}\,,\qquad
    a_n=\frac{\left(-1\right)^{n}\left(n+3\right)!}{n!\,3!}\,, \nn
    &&b_n = \frac{\left(n+2\right)\left(n+3\right)a_{n-1}}{4n}=-\,\frac{\left(-1\right)^{n}\left(n+2\right)\left(n+3\right)!}{n!\,4!}\,.
\eea
Using these relations, the above series can be summed up to the expression
\bea
    L^{(+4)}=\frac{\hat{Y}^{(+2)}\hat{Y}^{(+2)}}{\left(1+c^{(-2)}\hat{Y}^{(+2)}\right)^4}
    +\frac{m^2}{2}\left(\theta^{(+)}\right)^4\left[1-\frac{1-c^{(-2)}\hat{Y}^{(+2)}}{\left(1+c^{(-2)}\hat{Y}^{(+2)}\right)^5}\right].
\eea
For calculating the component Lagrangians, we use the following identities (up to total harmonic derivatives):
\bea
    c^{(-2)}\hat{y}^{(+2)}&=&\frac{1}{12}\,c^{IJ}\hat{y}_{IJ}\,,\nn
    \left(c^{(-2)}\hat{y}^{(+2)}\right)^{2}&=&\frac{1}{80}\left[\left(c^{IJ}\hat{y}_{IJ}\right)^{2}-\frac{2}{3}\,\hat{y}^{IJ}\hat{y}_{IJ}\right],\nn
    \left(c^{(-2)}\hat{y}^{(+2)}\right)^{3}&=&\frac{1}{400}\left[\left(c^{IJ}\hat{y}_{IJ}\right)^{3}-\frac{3}{2}\,\hat{y}^{KL}\hat{y}_{KL}\,c^{IJ}y_{IJ}\right],\nn
    \left(c^{(-2)}\hat{y}^{(+2)}\right)^{2n}&=&\frac{12\left(2n\right)!\left(2n\right)!}{2^{2n}\left(2n+2\right)!\left(2n+3\right)!}\nn
    &&\times\,\sum_{k=0}^{n}\frac{\left(-1\right)^k\left(2n-k+1\right)!}{k!\left(2n-2k\right)!}\left(\hat{y}^{KL}\hat{y}_{KL}\right)^k\left(c^{IJ}\hat{y}_{IJ}\right)^{2n-2k},\nn
    \left(c^{(-2)}\hat{y}^{(+2)}\right)^{2n+1}&=&\frac{12\left(2n+1\right)!\left(2n+1\right)!}{2^{2n+1}\left(2n+3\right)!\left(2n+4\right)!}\nn
    &&\times\,\sum_{k=0}^{n}\frac{\left(-1\right)^k\left(2n-k+2\right)!}{k!\left(2n-2k+1\right)!}\left(\hat{y}^{KL}\hat{y}_{KL}\right)^k\left(c^{IJ}\hat{y}_{IJ}\right)^{2n-2k+1}.
\eea
Using them, one obtains
\bea
    \frac{\partial^2}{\partial \hat{y}^{(+2)}\partial \hat{y}^{(+2)}}\left[\frac{\hat{y}^{(+2)}\hat{y}^{(+2)}}{\left(1+c^{(-2)}\hat{y}^{(+2)}\right)^4}\right]&=&\sum_{n=0}^{\infty}
    \left(n+1\right)\left(n+2\right)a_n\left(c^{(-2)}\hat{y}^{(+2)}\right)^{n}\nn
    &=&2\left(1+\frac{c^{IJ}\hat{y}_{IJ}}{2}+\frac{\hat{y}^{IJ}\hat{y}_{IJ}}{4}\right)^{-2}\nn
    &=&8\left[\frac{1}{2}\,y^{IJ}y_{IJ}\right]^{-2},\nn
    \sum_{n=1}^{\infty} b_n\left(c^{(-2)}\hat{Y}^{(+2)}\right)^{n}&=&-\,\frac{1}{2}\left(1+\frac{c^{IJ}\hat{y}_{IJ}}{2}+\frac{\hat{y}^{IJ}\hat{y}_{IJ}}{4}\right)^{-1}\nn
    &=&-\left[\frac{1}{2}\,y^{IJ}y_{IJ}\right]^{-1}.
\eea

\section{The multiplet ${\bf (6,8,2)}$}\label{682}
As was already mentioned, the multiplets ${\bf (6,8,2)}$ and ${\bf (8,8,0)}$ are related to each other by the substitution $i\dot{\phi}=D$. The same substitution is admissible in the Lagrangian \eqref{so6}. Then, the ${\rm SU}(4|1)$ invariant action \eqref{L+4act} is given by
\bea
    {\cal L}_{\bf (6,8,2)} &=& g_2\left[\frac{1}{2}\,\dot{y}^{IJ}\dot{y}_{IJ}+ \frac{i}{2}\left(\chi^K\dot{\bar{\chi}}_K-\dot{\chi}^K\bar{\chi}_K\right) + D\bar{D} - \frac{m}{4}\,\chi^K\bar{\chi}_K-\frac{m^2}{8}\,y^{IJ}y_{IJ}\right]\nn
    &&+\,\frac{1}{\sqrt{2}}\,\bar{D}\,\partial_{IJ}g_2\,\chi^I\chi^J - \frac{1}{\sqrt{2}}\,D\,\partial^{IJ}g_2\,\bar{\chi}_I\bar{\chi}_J +i\left(\dot{y}_{IK}\,\partial^{JK}g_2-\dot{y}^{JK}\,\partial_{IK}g_2\right)\chi^I\bar{\chi}_J\nn
    &&-\,\frac{1}{2}\,\partial_{IJ}\partial^{KL}g_2\,\chi^I\chi^J\bar{\chi}_K\bar{\chi}_L\,,\qquad g_2 = \left[\frac{1}{2}\,y^{IJ}y_{IJ}\right]^{-2}.\label{L682}
\eea
One can obtain a superconformal Lagrangian by making the following  substitutions of the component fields in \eqref{L682}
\bea
    D \rightarrow D\,e^{imt/2},\qquad \bar{D} \rightarrow \bar{D}\,e^{-imt/2},\qquad
    \chi^I \rightarrow \chi^I\, e^{imt/4},\qquad \bar{\chi}_I \rightarrow \bar{\chi}_I\,e^{-imt/4}.
\eea
The resulting  Lagrangian
\bea
    {\cal L}_{\rm conf} &=& g_2\left[\frac{1}{2}\,\dot{y}^{IJ}\dot{y}_{IJ}+ \frac{i}{2}\left(\chi^K\dot{\bar{\chi}}_K-\dot{\chi}^K\bar{\chi}_K\right) + D\bar{D}-\frac{m^2}{8}\,y^{IJ}y_{IJ}\right]\nn
    &&+\,\frac{1}{\sqrt{2}}\,\bar{D}\,\partial_{IJ}g_2\,\chi^I\chi^J - \frac{1}{\sqrt{2}}\,D\,\partial^{IJ}g_2\,\bar{\chi}_I\bar{\chi}_J +i\left(\dot{y}_{IK}\,\partial^{JK}g_2-\dot{y}^{JK}\,\partial_{IK}g_2\right)\chi^I\bar{\chi}_J\nn
    &&-\,\frac{1}{2}\,\partial_{IJ}\partial^{KL}g_2\,\chi^I\chi^J\bar{\chi}_K\bar{\chi}_L \label{L682conf}
\eea
is an even function of $m$ and so is superconformal. Its form is typical for Lagrangians with the trigonometric realizations of superconformal groups.

It can be shown that the superconformal group of \eqref{L682conf} is ${\rm SU}(4|1,1)$.
On the one hand, the transformations \eqref{tr682} become
\bea
    &&\delta D = -\,\sqrt{2}\,\epsilon_{I}\left(i\dot{\chi}^{I}-m\,\chi^I\right)e^{imt/2},\qquad \delta \bar{D} = -\,\sqrt{2}\,\bar{\epsilon}^{I}\left(i\dot{\bar{\chi}}_{I}+m\,\bar{\chi}_{I}\right)e^{-imt/2},\nn
    &&\delta y^{IJ} = -\,2\,\bar{\epsilon}^{[I}\chi^{J]} e^{-imt/2}+ \varepsilon^{IJKL}\epsilon_{K}\bar{\chi}_{L}\,e^{imt/2},\nn
    &&\delta \chi^I = \sqrt{2}\,\bar{\epsilon}^I B\, e^{-imt/2} - 2\,\epsilon_J\left(i\dot{y}^{IJ}-\frac{m}{2}\,y^{IJ}\right)e^{imt/2},\nn
    &&\delta \bar{\chi}_I = \sqrt{2}\,\epsilon_I \bar{B}\,e^{imt/2} + 2\,\bar{\epsilon}^J\left(i\dot{y}_{IJ}+\frac{m}{2}\,y_{IJ}\right)e^{-imt/2}, \label{tr682conf1}
\eea
and leave the Lagrangian \eqref{L682conf} invariant.
On the other hand,
new ${\rm SU}(4|1)$ transformations leaving invariant \eqref{L682conf} are defined by replacing $m\rightarrow-m$ in \eqref{tr682conf1}\,:
\bea
    &&\delta D = -\,\sqrt{2}\,\eta_{I}\left(i\dot{\chi}^{I}+m\,\chi^I\right)e^{-imt/2},\qquad \delta \bar{D} = -\,\sqrt{2}\,\bar{\eta}^{I}\left(i\dot{\bar{\chi}}_{I}-m\,\bar{\chi}_{I}\right)e^{imt/2},\nn
    &&\delta y^{IJ} = -\,2\,\bar{\eta}^{[I}\chi^{J]} e^{imt/2}+ \varepsilon^{IJKL}\eta_{K}\bar{\chi}_{L}\,e^{-imt/2},\nn
    &&\delta \chi^I = \sqrt{2}\,\bar{\eta}^I B\, e^{imt/2} - 2\,\eta_J\left(i\dot{y}^{IJ}+\frac{m}{2}\,y^{IJ}\right)e^{-imt/2},\nn
    &&\delta \bar{\chi}_I = \sqrt{2}\,\eta_I \bar{B}\,e^{-imt/2} + 2\,\bar{\eta}^J\left(i\dot{y}_{IJ}-\frac{m}{2}\,y_{IJ}\right)e^{imt/2}. \label{tr682conf2}
\eea
Introducing the conformal Hamiltonian as
\be
{\cal H}^{\prime}_{\rm conf} = {\cal H}+\frac{m}{2}\,F,
\ee
the superconformal algebra $su(4|1,1)$ amounts to the following set of (anti)commutators
\bea
    &&\left\lbrace Q^{I}, \bar{Q}_{J}\right\rbrace = 2\delta^{I}_{J}{\cal H}^{\prime}_{\rm conf}+2m\,L^{I}_{J} - m\,\delta^{I}_{J}F,\nn
    &&\left\lbrace S^{I}, \bar{S}_{J}\right\rbrace = 2\delta^{I}_{J}{\cal H}^{\prime}_{\rm conf}-2m\,L^{I}_{J} + m\,\delta^{I}_{J}F,\nn
    &&\left\lbrace Q^{I}, \bar{S}_{J}\right\rbrace = 2\delta^{I}_{J}T^{\prime},\qquad
    \left\lbrace S^{I}, \bar{Q}_{J}\right\rbrace = 2\delta^{I}_{J}\bar{T}^{\prime},
\eea
\bea
    &&\left[L^I_J,  L^K_L\right]
    = \delta^K_J L^I_L - \delta^I_L L^K_J,\nn
    &&\left[T^{\prime}, \bar{T}^{\prime}\right]=2m\,{\cal H}^{\prime}_{\rm conf}\,,\qquad
    \left[{\cal H}^{\prime}_{\rm conf}, T^{\prime}\right] = -\,m\,T^{\prime},\qquad
    \left[{\cal H}^{\prime}_{\rm conf}, \bar{T}^{\prime}\right] = m\,\bar{T}^{\prime},
\eea
\bea
    &&\left[L^I_J, Q^{K}\right]
    = \delta^K_J Q^{I} - \frac{1}{4}\,\delta^I_J Q^{K} ,\qquad
    \left[L^I_J, \bar{Q}_{L}\right] = \frac{1}{4}\,\delta^I_J\bar{Q}_{L}-\delta^I_L\bar{Q}_{J}\,,\nn
    &&\left[L^I_J, S^{K}\right]
    = \delta^K_J S^{I} - \frac{1}{4}\,\delta^I_J S^{K} ,\qquad
    \left[L^I_J, \bar{S}_{L}\right] = \frac{1}{4}\,\delta^I_J\bar{S}_{L}-\delta^I_L\bar{S}_{J}\,,\nn
    &&\left[F, Q^{I}\right]=\frac{1}{2}\,Q^{I},\quad
    \left[F, \bar{Q}_{J}\right]=-\,\frac{1}{2}\,\bar{Q}_{J}\,,\quad
    \left[F, S^{I}\right]=\frac{1}{2}\,S^{I},\quad
    \left[F, \bar{S}_{J}\right]=-\,\frac{1}{2}\,\bar{S}_{J}\,,\nn
    &&\left[{\cal H}^{\prime}_{\rm conf}, Q^{I}\right]=-\,\frac{m}{2}\,Q^{I},\qquad
    \left[{\cal H}^{\prime}_{\rm conf}, \bar{Q}_{J}\right]=\frac{m}{2}\,\bar{Q}_{J}\,,\nn
    &&\left[{\cal H}^{\prime}_{\rm conf}, S^{I}\right]=\frac{m}{2}\,S^{I},\qquad
    \left[{\cal H}^{\prime}_{\rm conf}, \bar{S}_{J}\right]=-\,\frac{m}{2}\,\bar{S}_{J}\,,\nn
    &&\left[T^{\prime},\bar{Q}_{J}\right]=m\,\bar{S}_{J}\,,\qquad
    \left[T^{\prime},S^I\right]=m\,Q^I,\nn
    &&\left[\bar{T}^{\prime},\bar{S}_{J}\right]=-\,m\,\bar{Q}_{J}\,,\qquad
    \qquad
    \left[\bar{T}^{\prime},Q^I\right]=-\,m\,S^I.
\eea
Note that the ``parabolic'' realization of the superconformal group SU$(4|1,1)$ on the undeformed multiplet $({\bf 6, 8, 2})$ was
given in \cite{KhTo}. The corresponding superconformal Lagrangian can be obtained as the $m{=}0$ limit of \eqref{L682conf}.

\section{The multiplet ${\bf (7,8,1)}$}\label{781}
The substitution $\sqrt{2}\,i\dot{\phi}=C-i\dot{x}$ in \eqref{tr880} gives ${\rm SU}(4|1)$ transformations of the multiplet ${\bf (7,8,1)}$\,:
\bea
    &&\delta C = -\,\epsilon_{I}\left(i\dot{\chi}^{I}-\frac{3m}{4}\,\chi^{I}\right)e^{3imt/4}-\bar{\epsilon}^{I}\left(i\dot{\bar{\chi}}_{I}+\frac{3m}{4}\,\bar{\chi}_{I}\right)e^{-3imt/4},\nn
    &&\delta x = \epsilon_{I}\chi^{I}e^{3imt/4}-\bar{\epsilon}^{I}\bar{\chi}_{I}\,e^{-3imt/4},\nn
    &&\delta y^{IJ} = -\,2\,\bar{\epsilon}^{[I}\chi^{J]} e^{-3imt/4}+ \varepsilon^{IJKL}\epsilon_{K}\bar{\chi}_{L}\,e^{3imt/4},\nn
    &&\delta \chi^I = \bar{\epsilon}^I\left(C-i\dot{x}\right) e^{-3imt/4} - 2\,\epsilon_J\left(i\dot{y}^{IJ}-\frac{m}{2}\,y^{IJ}\right)e^{3imt/4},\nn
    &&\delta \bar{\chi}_I = \epsilon_I\left(C+i\dot{x}\right)e^{3imt/4} + 2\,\bar{\epsilon}^J\left(i\dot{y}_{IJ}+\frac{m}{2}\,y_{IJ}\right)e^{-3imt/4}. \label{tr781}
\eea
The same substitution is admissible in the Lagrangian \eqref{so6}. In this way we obtain the  ${\rm SU}(4|1)$ invariant Lagrangian for the multiplet ${\bf (7,8,1)}$\,:
\bea
    {\cal L}_{\bf (7,8,1)} &=& g_2\left[\frac{\dot{x}^2}{2} + \frac{1}{2}\,\dot{y}^{IJ}\dot{y}_{IJ}+ \frac{i}{2}\left(\chi^K\dot{\bar{\chi}}_K-\dot{\chi}^K\bar{\chi}_K\right) + \frac{C^2}{2} - \frac{m}{4}\,\chi^K\bar{\chi}_K-\frac{m^2}{8}\,y^{IJ}y_{IJ}\right]\nn
    &&+\,\frac{C}{2}\left(\partial_{IJ}g_2\,\chi^I\chi^J - \partial^{IJ}g_2\,\bar{\chi}_I\bar{\chi}_J\right) + \frac{i\dot{x}}{2}\left(\partial_{IJ}g_2\,\chi^I\chi^J + \partial^{IJ}g_2\,\bar{\chi}_I\bar{\chi}_J\right)\nn
    &&+\,i\left(\dot{y}_{IK}\,\partial^{JK}g_2-\dot{y}^{JK}\,\partial_{IK}g_2\right)\chi^I\bar{\chi}_J
    -\frac{1}{2}\,\partial_{IJ}\partial^{KL}g_2\,\chi^I\chi^J\bar{\chi}_K\bar{\chi}_L\,,\nn
    && g_2 = \left[\frac{1}{2}\,y^{IJ}y_{IJ}\right]^{-2}.\label{L781}
\eea
The corresponding action is not invariant with respect to the superconformal group F(4) inherent to the multiplet ${\bf (7,8,1)}$ \cite{FT2}
because \eqref{L781}  cannot be brought to a form in which it depends only on  $m^2$. On the other hand, since F(4) includes SU$(4|1)$ as a subgroup,
we  expect the existence of an alternative, F(4) superconformal, action for the SU$(4|1)$ multiplet ${\bf (7,8,1)}$.
At the component level, such a system has recently been constructed, without giving an action however~\cite{AiKuTo}.

\section{Superconformal algebra $osp(8|2)$}\label{osp82}
Superconformal algebra $osp(8|2)$ is given by the following non-vanishing (anti)commutators:
\bea
    &&\left\lbrace Q^{I}, \bar{Q}_{J}\right\rbrace = 2\delta^{I}_{J}{\cal H}_{\rm conf}+2m\,L^{I}_{J} + m\,\delta^{I}_{J}F,\nn
    &&\left\lbrace S^{I}, \bar{S}_{J}\right\rbrace = 2\delta^{I}_{J}{\cal H}_{\rm conf}-2m\,L^{I}_{J} - m\,\delta^{I}_{J}F,\nn
    &&\left\lbrace Q^{I}, S^{J}\right\rbrace = 2m\,F^{IJ},\qquad
    \left\lbrace \bar{Q}_{I}, \bar{S}_{J}\right\rbrace = 2m\,\bar{F}_{IJ}\,,\nn
    &&\left\lbrace Q^{I}, \bar{S}_{J}\right\rbrace = 2\delta^{I}_{J}T,\qquad
    \left\lbrace S^{I}, \bar{Q}_{J}\right\rbrace = 2\delta^{I}_{J}\bar{T},\label{conf_algebra_osp82_ff}
\eea
\bea
    &&\left[L^I_J,  L^K_L\right]
    = \delta^K_J L^I_L - \delta^I_L L^K_J,\nn
    &&\left[L^I_J, F^{KL}\right]
    = \delta^K_J F^{IL}+\delta^L_J F^{KI} - \frac{1}{2}\,\delta^I_J F^{KL} ,\nn
    &&\left[L^I_J, \bar{F}_{KL}\right] = \frac{1}{2}\,\delta^I_J\bar{F}_{KL}-\delta^I_K\bar{F}_{JL}-\delta^I_L\bar{F}_{KJ}\,,\nn
    &&\left[F^{IJ}, \bar{F}_{KL}\right] = \delta^I_K L^J_{L} - \delta^I_L L^J_{K}+ \delta^J_{L} L^I_K - \delta^J_{K} L^I_L + \left(\delta^I_K\delta^J_L - \delta^I_L\delta^J_K\right) F,\nn
    &&\left[F, F^{IJ}\right]=F^{IJ},\qquad
    \left[F, \bar{F}_{IJ}\right]=-\,\bar{F}_{IJ}\,,\nn
    &&\left[T, \bar{T}\right]=4m\,{\cal H}_{\rm conf}\,,\qquad
    \left[{\cal H}_{\rm conf}, T\right] = -\,2m\,T,\qquad
    \left[{\cal H}_{\rm conf}, \bar{T}\right] = 2m\,\bar{T},
    \label{conf_algebra_osp82_bb}
\eea
\bea
    &&\left[L^I_J, Q^{K}\right]
    = \delta^K_J Q^{I} - \frac{1}{4}\,\delta^I_J Q^{K} ,\qquad
    \left[L^I_J, \bar{Q}_{L}\right] = \frac{1}{4}\,\delta^I_J\bar{Q}_{L}-\delta^I_L\bar{Q}_{J}\,,\nn
    &&\left[L^I_J, S^{K}\right]
    = \delta^K_J S^{I} - \frac{1}{4}\,\delta^I_J S^{K} ,\qquad
    \left[L^I_J, \bar{S}_{L}\right] = \frac{1}{4}\,\delta^I_J\bar{S}_{L}-\delta^I_L\bar{S}_{J}\,,\nn
    &&\left[\bar{F}_{IJ}, Q^{K}\right]=\delta^K_I\bar{S}_{J}-\delta^K_J\bar{S}_{I}\,,\qquad
    \left[F^{IJ}, \bar{Q}_{L}\right]=\delta^J_L S^I-\delta^I_L S^J,\nn
    &&\left[\bar{F}_{IJ}, S^{K}\right]=\delta^K_I\bar{Q}_{J}-\delta^K_J\bar{Q}_{I}\,,\qquad
    \left[F^{IJ}, \bar{S}_{L}\right]=\delta^J_L Q^I-\delta^I_L Q^J,\nn
    &&\left[F, Q^{I}\right]=\frac{1}{2}\,Q^{I},\qquad
    \left[F, \bar{Q}_{J}\right]=-\,\frac{1}{2}\,\bar{Q}_{J}\,,\nn
    &&\left[F, S^{I}\right]=\frac{1}{2}\,S^{I},\qquad
    \left[F, \bar{S}_{J}\right]=-\,\frac{1}{2}\,\bar{S}_{J}\,,\nn
    &&\left[{\cal H}_{\rm conf}, Q^{I}\right]=-\,m\,Q^{I},\qquad
    \left[{\cal H}_{\rm conf}, \bar{Q}_{J}\right]=m\,\bar{Q}_{J}\,,\nn
    &&\left[{\cal H}_{\rm conf}, S^{I}\right]=m\,S^{I},\qquad
    \left[{\cal H}_{\rm conf}, \bar{S}_{J}\right]=-\,m\,\bar{S}_{J}\,,\nn
    &&\left[T,\bar{Q}_{J}\right]=2m\,\bar{S}_{J}\,,\qquad
    \left[T,S^I\right]=2m\,Q^I,\nn
    &&\left[\bar{T},\bar{S}_{J}\right]=-\,2m\,\bar{Q}_{J}\,,\qquad
    \qquad
    \left[\bar{T},Q^I\right]=-\,2m\,S^I.
\label{conf_algebra_osp82_bf}
\eea
The supercharges $Q^I$, $\bar{Q}_J$ together with the generators $L^I_J$ and $H = {\cal H}_{\rm conf} + \frac{m}{2}F$ form the subalgebra $su(4|1)\; +\!\!\!\!\!\!\supset u(1)$
in  $osp(8|2)$, with $F$ being an additional external $R$-symmetry ${\rm U}(1)$ generator.
The second set of ${\rm SU}(4|1)$ supercharges $S^I$, $\bar{S}_J$ extends this subalgebra to the full superconformal algebra $osp(8|2)$.
The latter involves twelve additional $R$-symmetry generators $F^{IJ}\equiv F^{[IJ]}$, $\bar{F}_{IJ}\equiv\bar{F}_{[IJ]}$ which, together with the ${\rm U}(4)$ generators $L^I_J, F$,
form the full $R$-symmetry algebra $o(8)$. Additional conformal generators are $\bar{T}$, $T$, such that  three bosonic generators ${\cal H}_{\rm conf}$\,, $\bar{T}$ and $T$
constitute the conformal $d{=}1$ subalgebra $o(2,1)$.

Actually, the parameter $m$ drops out from the superconformal algebra after performing redefinitions similar to those made in \cite{ISTconf}
for the case of the ${\cal N}\,{=}\,4, d\,{=}\,1$ superconformal algebra $D(2,1;\alpha)\,$.

\end{document}